\documentclass{cpbtex}

\usepackage{color}
\usepackage[utf8]{inputenc}
\usepackage{tabularx}
\usepackage{amsmath}
\usepackage{amssymb}
\usepackage{longtable}
\usepackage{siunitx}
\usepackage[sort&compress,numbers]{natbib}	
\usepackage{doi}
\usepackage{hyperref}
\usepackage{mathptmx}       
\usepackage{helvet}         
\usepackage{courier}        
\usepackage{type1cm}        
\usepackage{makeidx}         
\usepackage{graphicx}        
\usepackage{multicol}        
\usepackage[bottom]{footmisc}
\usepackage{hyperref}
\hypersetup{
  pdfnewwindow=true, colorlinks=true,
  linkcolor=blue, anchorcolor=blue,
  citecolor=blue, filecolor=blue,
  menucolor=blue, urlcolor=blue}

\providecommand{\bk}{\ensuremath{{\boldsymbol k}}}
\providecommand{\kp}{\ensuremath{{\boldsymbol k}\cdot{\boldsymbol p}}}

\providecommand{\inassb}{I\lowercase{n}A\lowercase{s}$_{1-x}$S\lowercase{b}$_x$}
\providecommand{\inassbh}{I\lowercase{n}A\lowercase{s}$_{0.5}$S\lowercase{b}$_{0.5}$}

\begin{document}
\title{Topology of Triple-Point Metals}

\author{Georg W. Winkler$^{1}$, \ Sobhit Singh$^{2}$,  \ and \ Alexey A. Soluyanov$^{3,4}$\thanks{Corresponding author. E-mail: asoluyan@physik.uzh.ch}\\
\small{$^{1}${Microsoft Quantum, Microsoft Station Q, University of California, Santa Barbara, California 93106-6105 USA}}\\   
\small{$^{2}${Department of Physics and Astronomy, Rutgers University, Piscataway, New Jersey 08854 USA}}\\ 
\small{$^{3}${Institute of Physics, University of Zurich, Winterthurerstrasse 190, 8057 Zurich, Switzerland}}\\
\small{$^{4}${Department of Physics, St. Petersburg State University, St. Petersburg, 199034 Russia }}} 

\maketitle

\begin{abstract}
We discuss and illustrate the appearance of topological fermions and bosons in triple-point metals, where a band crossing of three electronic bands occurs close to the Fermi level. Topological bosons appear in the phonon spectrum of certain triple-point metals, depending on the mass of atoms that form the binary triple-point metal. We first provide a classification of possible triple-point electronic topological phases possible in crystalline compounds and discuss the consequences of these topological phases, seen in Fermi arcs, topological Lifshitz transitions and transport anomalies. Then we show how the topological phase of phonon modes can be extracted and proven for relevant compounds. Finally, we show how the interplay of electronic and phononic topologies in triple-point metals puts these \textit{metallic} materials into the list of the most efficient \textit{metallic} thermoelectrics known to date.
\end{abstract}




\section{Introduction}

\begin{figure}
  \includegraphics[width = \linewidth]{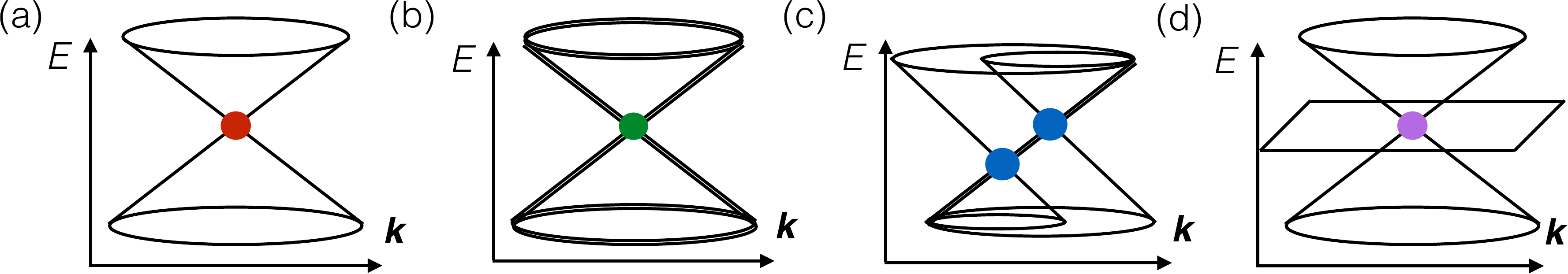}
  \caption{(a) Weyl points are twofold degenerate band crossings of
    two singly degenerate bands. (b) Dirac points are fourfold
    degenerate band crossings of two twofold degenerate bands. (c)
    Triple-points are threefold degenerate band crossings of a singly
    and a doubly degenerate band. Here a pair is shown. (d) A spin-1
    generalization of a Weyl fermion~\cite{bernevig_triple_2016}.}
  \label{fig:weyl_dirac_triple}
\end{figure}

Materials with non-trivial band structure topology, apart from
possible technological applications, provide a test ground for the
concepts of fundamental physics theories in relatively cheap condensed
matter experiments. For example, the recent discovery of Weyl
semimetals in TaAs materials class~\cite{weng2015weyl, hasan_nat_comm,
  Xu_TaAs, wang_TaAs, Lv_TaAs} provided materials, where two bands
cross linearly at isolated points in momentum space as illustrated in
Fig.~\ref{fig:weyl_dirac_triple}~(a), called Weyl points
(WPs)~\cite{Vishwanath}. These WPs occur close to the Fermi level, and
hence the low energy excitations in these metals are described by the
Weyl equation of the relativistic quantum field theory, thus allowing
for experimental studies of Weyl fermions, examples of which in
high-energy physics are still lacking.

Another example of a topological material hosting a quasiparticle
analogue of an elementary particle is that of Dirac
semimetals~\cite{wang_dirac_2012,wang_cd3as2,cd3as2_experiment,Xu_dirac_arpes}. These
are centrosymmetric non-magnetic materials that host Dirac points
(DPs) -- points of linear crossing of two doubly degenerate bands in
momentum space, see Fig.~\ref{fig:weyl_dirac_triple}~(b). When DPs are
located close to the Fermi level, the low energy excitations of the
hosting metal are described by the Dirac equation, and thus become
direct analogues of Dirac electrons in high-energy theories.

More recently, it was shown that a variety of possible symmetries
realized in solids also allows for the existence of topological
quasiparticle excitations, which do not have direct analogues in the
Standard Model~\cite{alexey_weyl, hourglass, kim2015dirac, nodal_chains, Kruthoff-PRX17, kane_newfermions, bernevig_triple_2016, burkov_balents, nodal_chains, Vishwanath-Parameswaran}, rendering
novel physical behavior to the hosting compounds. Classification and
description of possible topologically protected quasiparticles in
solids, along with the identification of material candidates, becomes
of major importance for the progress in materials science and
technology, as well as condensed matter theory in general.

The subject of this review is the so-called triple-point (TP)
fermionic quasiparticle, which is a crossing of a singly and a doubly
degenerate band, see
Fig.~\ref{fig:weyl_dirac_triple}~(c)~\cite{winkler2016topological,
  winkler2016triple, xi_dai_triple}. Apart from the TPs another class
of topologically distinct threefold degenerate band crossing exists
representing a spin-1 generalization of a Weyl fermion, see
Fig.~\ref{fig:weyl_dirac_triple}~(d). While the TPs occur as
accidental degeneracies in both symmorphic and non-symmorphic crystal
structures, the latter is limited to high symmetry points in certain
non-symmorphic space groups, containing symmetries combined of a point
group symmetry operation followed by translation by a fraction of the
primitive unit cell vector~\cite{bernevig_triple_2016}.  While the TPs
described here can be also found in non-symmorphic space groups, the
symmetry conditions for their appearance coincides in such cases with
those of symmorphic space groups. The spin-1 fermion is furthermore
characterized by a nontrivial Chern number, which is ill defined for
the TPs due to the doubly degenerate band. The nontrivial topology of
TPs, on the other hand, is manifested by topologically protected nodal
lines and a $\mathbb Z_2$ topological classification applicable for
pairs of TPs.

\section{Classification of Triple-Points}

The realization of a symmorphic TP at a momentum $\boldsymbol{k}$ in
the Brillouin zone (BZ) of a crystal structure requires the little
group of $\boldsymbol{k}$ to contain both one- and two-dimensional
double group representations, since both singly and doubly degenerate
bands need to be present for the formation of the TP. Only the point
group $C_{3v}$ satisfies these criteria, thus TPs appear on
high-symmetry lines in the BZ with the little group $C_{3v}$. The
elements of $C_{3v}$ are 3-fold rotation $C_3$ and 3 mirrors
$\sigma_v$, containing the $C_3$ axis, rotated by 120 degrees relative
to each
other~\cite{koster1963properties,bilbao,bradley2010mathematical}. One
notable exception from this rule is given by space group 174
($C^1_{3h}$), where the interplay of time-reversal and mirror symmetry
on a $C_3$-symmetric line also allows for both one- and
two-dimensional double group representations.

Using the above symmetry criterion all space groups that can host TP
fermions on a line are identified in Tab.~\ref{table:summary}. Note,
that the little group on the high-symmetry axis of the type-B TP
topological metals (TPTMs) is exactly $C_{3v}$, while for type-A TPTMs
it is supplemented by an additional anti-unitary symmetry. This
symmetry is the product of a mirror plane $\sigma_h$, orthogonal to
the $C_3$-axis and time-reversal (TR). Its presence preserves the
existence of doubly- and singly-degenerate representations, and,
hence, allows for the existence of TPs. In our consideration we also
included non-symmorphic space groups, such that the TP crossing
includes the same irreducible representations as found in the
symmorphic space groups.
\begin{table}
  \caption{Space groups allowing for TPs of different types with
    time-reversal symmetry. The points can appear on high-symmetry
    lines in the Brillouin zone: $\Delta=(0,0,\alpha)$, $\Lambda=(\alpha,\alpha,\alpha)$,
    $P=(-1/3,2/3,\alpha)$ (hexagonal lattice),
    $P=(1/2-\alpha,1/2-\alpha,-1/2-\alpha)$ (rhombohedral lattice) and
    $F=(1/4+\alpha,1/4-3\alpha,1/4+\alpha)$. Triple-points appear in
    trigonal, hexagonal and cubic space groups. Note that the table
    contains also non-symmorphic space groups but the TPs exist on
    lines where the non-symmorphicity does not change the irreducible
    representations, thus they are identical to the TPs found in
    symmorphic space groups. The case of the non-symmorphic space
    group 220 has also been treated in
    Ref.~\cite{bernevig_triple_2016}. We note that, in addition, the
    groups 162-167, 191-194 and 221-230 admit type-B TPs provided
    time-reversal symmetry is broken in a way preserving $C_{3v}$
    representations on a line in the Brillouin
    zone.}\label{table:summary}
  \begin{tabular}{p{1.5cm}p{2.95cm}p{4.85cm}p{3.8cm}}
   \\  \hline\noalign{\smallskip}
    TP type & $\Gamma$-A ($\Delta$) or $\Gamma$-P$_2$ ($\Lambda$) &
                                                                    K-H
                                                                    ($P$)
                                                                    or
                                                                    P$_0$-T
                                                                    ($P$)
    & $\Gamma$-L/R/P ($\Lambda$) or P-H ($F$)\\
       \hline\noalign{\smallskip}
    type-A & 174, 187-190 &  & \\
    \hline\noalign{\smallskip}
    type-B & 156-161  & 157, 159-161, 183-186, 189-190 & 215-220 \\
    \hline\noalign{\smallskip}
  \end{tabular}
\end{table}

\begin{figure}
  \includegraphics[width = \linewidth]{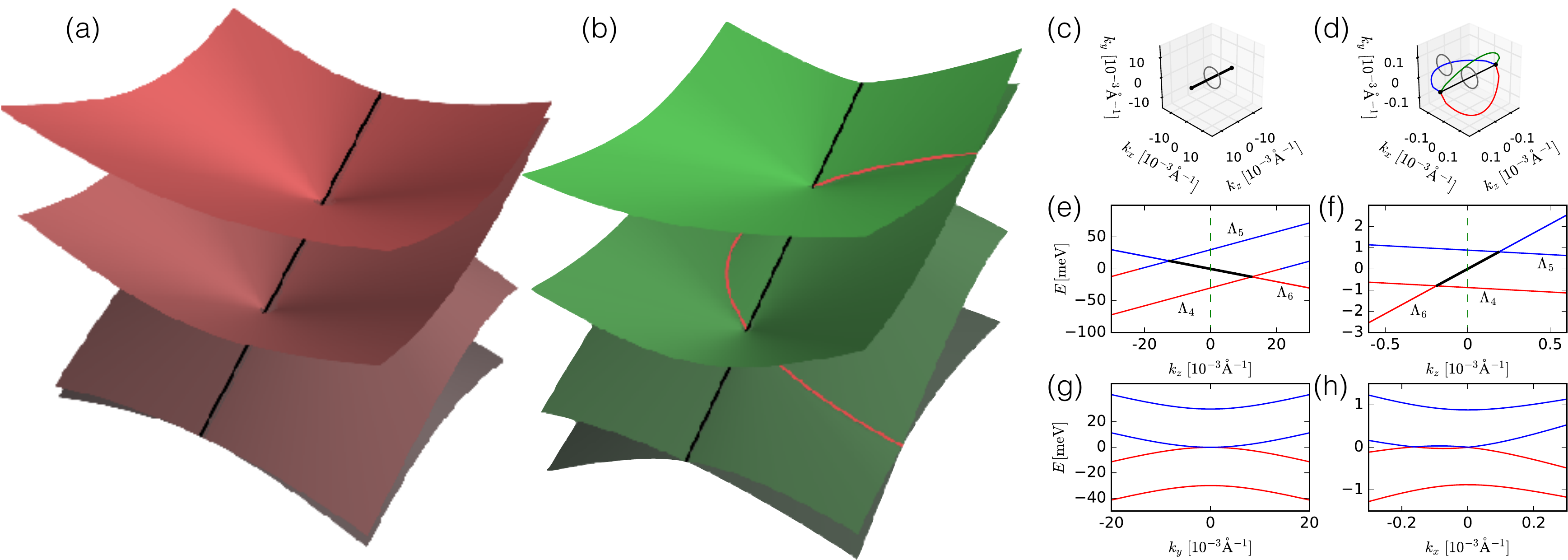}
  \caption{ Two types of fermionic triple-point quasiparticles. (a) and (c)
    type-A triple-points are connected by a single nodal line, where
    conduction and valence bands are degenerate (shown in black). (b)
    and (d) type-B triple-points are accompanied by four such nodal
    lines, shown in black, green, blue and red. The latter three occur
    in the mirror-symmetric planes in momentum space. The grey circles
    in (c) and (d) indicate paths for the Berry phase
    calculation. (e)((f)) Band structure around a type-A (type-B)
    triple-point along the $C_3$ axis. Here $\Lambda_6$ represents the
    double degenerate band (double representation of $C_{3v}$), while
    $\Lambda_{4,5}$ correspond to two one-dimensional representations.
    The black lines in (e) and (f) mark the region of the band
    structure that produces the nodal lines shown in black in panels
    (c) and (d). (g)((h)) Band structure around a type-A (type-B)
    triple-point in a mirror symmetric plane orthogonal to $k_z$. The
    dashed green lines in (e) and (f) mark the momentum $k_z$ used in
    panels (g) and (h). Red (blue) color in panels (e-h) corresponds
    to occupied (unoccupied) bands assuming the Fermi level is exactly
    between the pair of triple-points. Figure partially reused from
    Ref.~\cite{winkler2016triple}}
  \label{fig:nodal_lines}
\end{figure}

The topological classification of TPs into type-A and type-B stems
from the different numbers of accompanying nodal lines, and also from
the fact that the nodal lines accompanying the two types of TPs are
topologically distinct. Due to the three vertical mirror planes, the
Berry phase $\varphi_{\rm B}$ accumulated by valence bands on any
mirror-symmetric path (shown in grey in
Fig.~\ref{fig:nodal_lines}(c-d)) enclosing the corresponding nodal
line is quantized to be either $0$ or
$\pi$~\cite{alexandradinata_inversion_2014,schnyder_line_nodes}. The
nodal line of type-A TPTMs has $\varphi_{\rm B}=0$, while all the
lines of type-B TPs have $\varphi_{\rm B}=\pi$.

These values are consistent with the band structure plots, shown in
Fig.~\ref{fig:nodal_lines}(e-h). In type-A TPTMs the crossing of
conduction and valence (occupied and unoccupied) bands occurs on a
high-symmetry line and is quadratic, while for the type-B phase this
quadratic touching point splits into two points, where the bands cross
linearly. The presence of nodal lines with nontrivial Berry phase, as
is the case for type-B, is generally associated with the appearance of
surface states~\cite{weng_drumhead,schnyder_line_nodes}. The merging
and subsequent annihilation of nodal lines is similar to the nexus
point discussed in the context of $^3$He-A and Bernal-stacked graphite
with neglected spin-orbit coupling
(SOC)~\cite{volovik_book,mikitik_sharlai,nexus1,nexus2}. We stress,
however, that the scenarios discussed in the present work take full
account of SOC.

Analogous to WPs~\cite{Vishwanath}, the minimal number of TPs in the
BZ is four for materials preserving time-reversal symmetry. A pair of
TPs located on a $C_{3v}$-symmetric line can be split into four WPs by
lowering the $C_{3v}$ symmetry to $C_3$ (breaking $\sigma_v$), which
can be achieved by a small Zeeman field parallel to the $C_3$ axis or
by an atomic distortion. Conversely, imposing inversion symmetry onto
the atomic structure makes the two TPs merge into a single DP. Hence,
the TPTMs can be viewed as an intermediate phase separating Dirac and
Weyl semimetals in materials with a $C_{3v}$-symmetric line in the BZ.

\section{Triple-Point Materials}
\label{sec:materials}
The possibility of new fermions in condensed matter systems sparked a
huge effort in the first-principles community to look for suitable
material candidates. While band structures with threefold degenerate
crossings appear to be comparatively rare in nature, using extensive
scans of material databases enabled by high-throughput calculations, a
decent amount of material candidates for TPTMs could still be
identified. Experimental investigations of some of these materials are
currently under way with some results already
published~\cite{WC_magnetotransport,WC_arpes,MoP1,MoP2,InAsSb_experiment}. In
Tab.~\ref{table:materials} we list all TPTMs that have been predicted
by first-principles methods up to the point of this writing.

\begin{table}
  \caption{Theoretically predicted TPTMs classified in type-A and
    type-B. Material candidates for which experimental evidence of TPs
    exist are marked by an asterisk.}
  \label{table:materials}
  \begin{tabular}{ p{1.5cm}p{14.0cm}}
    \hline\noalign{\smallskip}
    TP type & Material\\
    \noalign{\smallskip}\hline\noalign{\smallskip}
    type-A &  MoC~\cite{winkler2016triple,MoC},
             WC$^*$~\cite{winkler2016triple,WC_magnetotransport,WC_arpes},
             WN~\cite{winkler2016triple},
             ZrTe~\cite{winkler2016triple,xi_dai_ZrTe}, 
             MoP$^*$~\cite{winkler2016triple,MoP1,MoP2},
             MoN~\cite{winkler2016triple},
             TaN~\cite{winkler2016triple,xi_dai_triple},
             NbN~\cite{winkler2016triple},
             NbS~\cite{winkler2016triple},
             TaS~\cite{TaS},
             TaSb~\cite{SinghPRM2018}\\
    \noalign{\smallskip}\hline\noalign{\smallskip}
    type-B &  tensile strained HgTe~\cite{zaheer_hgte},
             CuPt-ordered
             InAsSb$^*$~\cite{winkler2016topological,InAsSb_experiment},
             CuAgSe~\cite{CuAgSe},
             LuPtBi~\cite{half_heusler},
             LuAuPb~\cite{half_heusler},
             YPtBi~\cite{half_heusler},
             GdPtBi~\cite{half_heusler},
             LuPtBi~\cite{half_heusler},
             LaPtBi~\cite{half_heusler}\\
    \noalign{\smallskip}\hline\noalign{\smallskip}
  \end{tabular}
\end{table}

The TPTMs have been predicted in three types of crystal structures:
type-A TPTMs are exclusively of a tungsten carbide (WC) structure and
type-B TPTMs either have cubic symmetry (HgTe, half-Heuslers) or are
obtained by breaking the cubic $T_d$ symmetry to the subgroup $C_{3v}$
(tensile strained HgTe, CuPt-ordered InAsSb).

\subsection{   Triple-points in Tungsten Carbide like crystal structure}
\begin{figure}[t]
  \includegraphics[width = \linewidth]{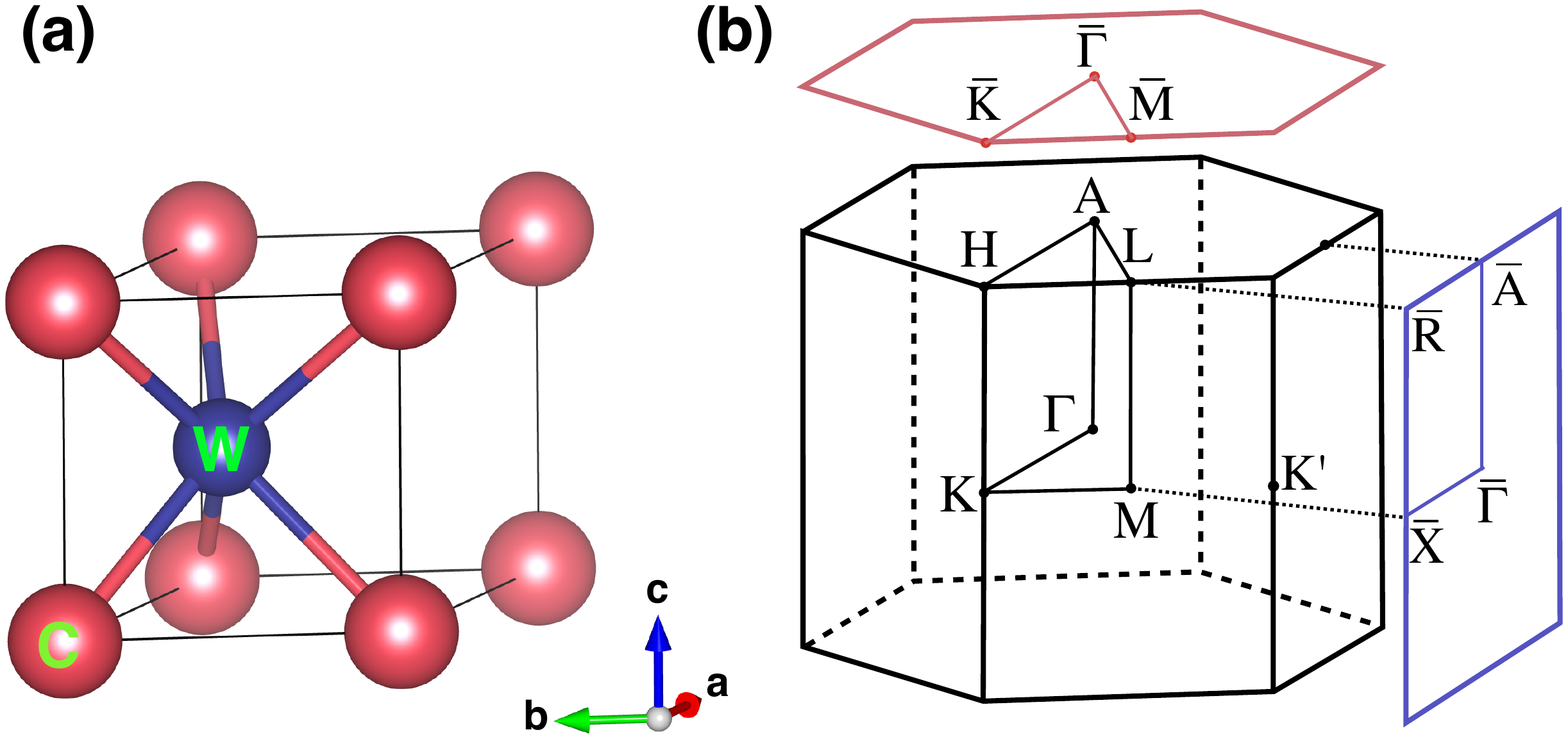}
  \caption{\label{fig:struct} (a) Primitive
    unit cell of WC-type structure. (b) The bulk BZ and (001) and
    (010) surface BZs. Figure reused from
    Ref.~\cite{winkler2016triple}.}

\end{figure}

The type-A TPTM phase has been identified in a family of two-element
metals AB (A=\{Zr, Nb, Mo, Ta, W\}, B=\{C, N, P, S, Te\}) listed in
the type-A row of Tab.~\ref{table:materials}. These materials have a
WC-type structure that belongs to space group $P\bar{6}m2$
($D_{3h}^1$=187). The primitive unit cell, shown in
Fig.~\ref{fig:struct}(a), consists of two atoms A and B at Wyckoff
positions 1a $(0,0,0)$ and 1d $(\frac{1}{3},\frac{2}{3},\frac{1}{2})$
respectively. The corresponding bulk BZ is shown in
Fig.~\ref{fig:struct}(b) along with the (001) and (010) surface BZs.

\begin{figure}
  \centerline{\includegraphics[width = \linewidth]{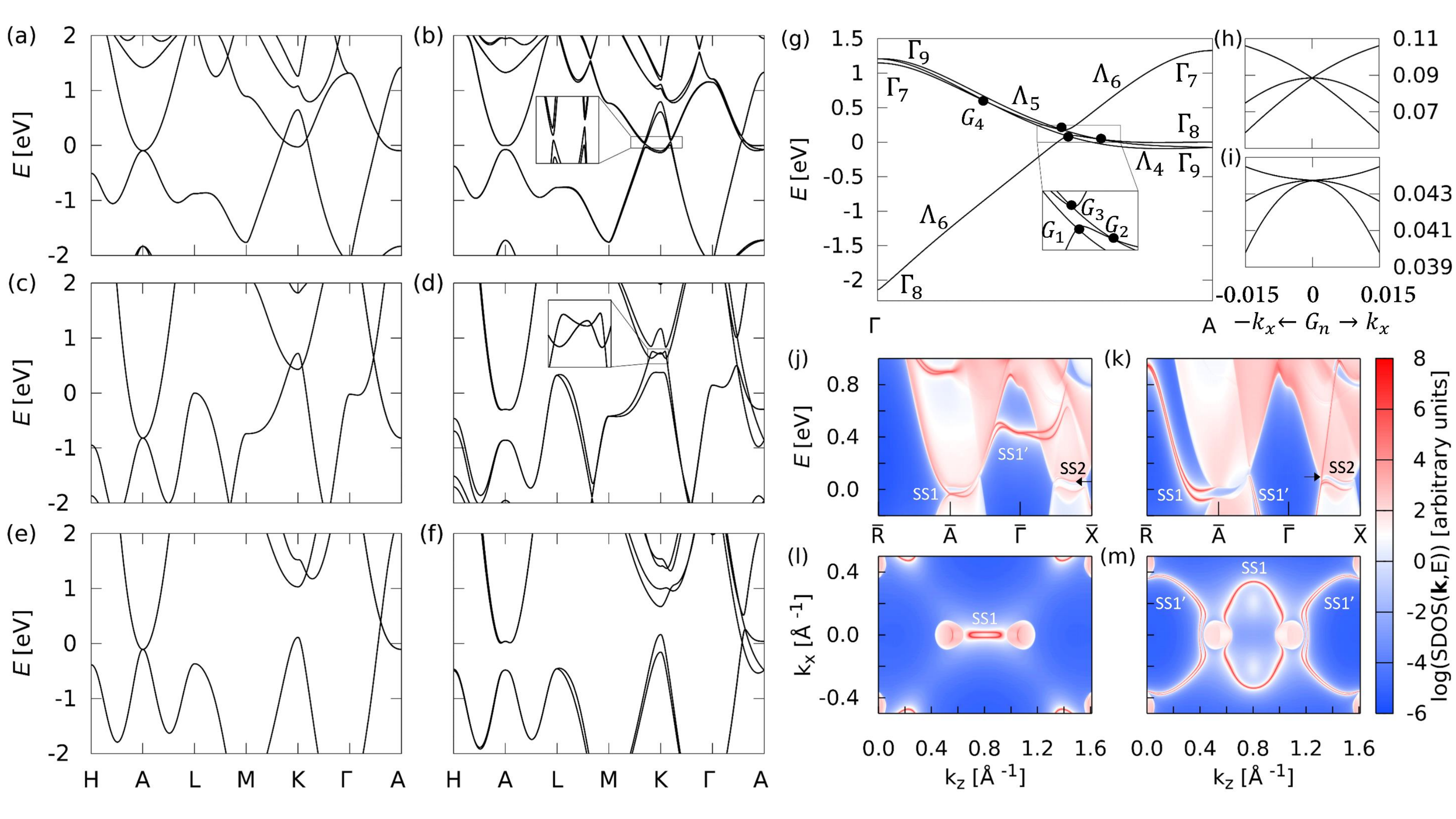}}
  \caption{\label{fig:fp} Band structure of ZrTe (a)((b)), WC
    (c)((d)), and TaN (e)((f)) without (with) SOC. The Fermi energy is
    set to 0 eV. (g) Band structure of ZrTe along the $\Gamma$-A
    line. Bands are labeled by their double group representations
    corresponding to $D_{3h}$ at $\Gamma$ and A points and $C_{3v}$ on
    the $\Gamma$-A line. (h)((i)) Band structures in the (100)
    direction with $k_z$ tuned to the TPs $G_1$ ($G_2$). (j) ((k))
    Projected surface density of states (SDOS) for the (010) surface
    of ZrTe with Zr (Te) termination. (l)((m)) The (010)-surface Fermi
    surface of ZrTe at $E=0$ eV for Zr (Te) termination. Figure reused
    from Ref.~\cite{winkler2016triple}.}
\end{figure}

Now we discuss the band structure obtained by \emph{ab initio}
simulations~\cite{winkler2016triple}. Fig.~\ref{fig:fp} shows the band
structures of ZrTe, WC and TaN, which will be used as representative
materials.

In the absence of SOC there is a band inversion at K and K' points in
ZrTe and WC (Fig.~\ref{fig:fp}(a) and (c)), resulting in a nodal ring
in the $k_z=0$ plane protected by $\sigma_h$, while this band
inversion is absent in TaN. The common feature of all materials is
that along the $C_{3v}$-symmetric $\Gamma$-A line there is a band
crossing of a singly and doubly degenerate bands due to the inversion
of the singly- ($\Lambda_1$) and the doubly-degenerate ($\Lambda_3$)
states at A. This crossing produces a single no-SOC TP, and it is this
feature that generates four TPs upon introducing SOC.

\begin{figure}
  \includegraphics[width = \linewidth]{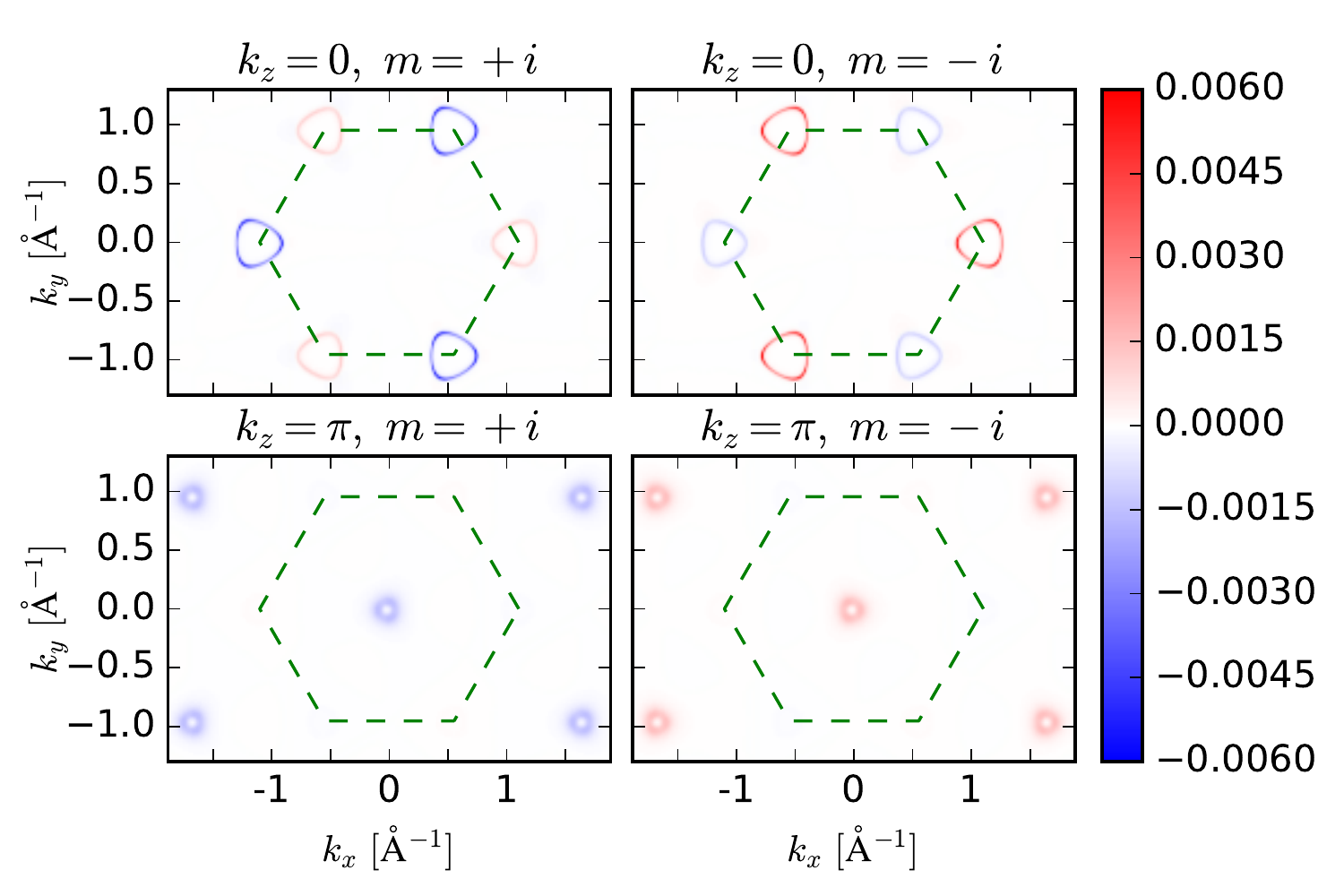}
  \caption{The Berry curvature for specific mirror eigenvalues on
    $\sigma_h$-mirror invariant planes in ZrTe. Figure reused from
    Ref.~\cite{winkler2016triple}.}
  \label{fig:mirror}
\end{figure}

All the considered materials have sizable SOC, which cannot be
neglected. Due to the lack of inversion symmetry the bands are
spin-split at generic momenta as shown in
Fig.~\ref{fig:fp}(b,d,f). One finds a band inversion along the H-A-L
line such that the A point acquires an inverted gap for all
materials. Consequently, the $k_z = \pi$ plane becomes an analogue of
a 2D quantum spin Hall insulator in all of the compounds. Topological
confirmation of the presence of band inversion is given by the
non-trivial values of the mirror Chern
numbers~\cite{mirror_chern_number} on the $\sigma_h$ planes, see
Fig.~\ref{fig:mirror}, which are $C_{\pm i} = \mp 1$, for the mirror
eigenvalues $\pm i$ respectively, in the $k_z=\pi$ plane for the
materials considered here~\cite{winkler2016triple}. In the $k_z=0$
plane one finds $C_{\pm i} = \mp 1$ for ZrTe, MoP and NbS and
$C_{\pm i} = 0$ for MoN, TaN and NbN. The materials MoC, WC and WN
have nodal lines in the $k_z=0$ plane and, therefore, the mirror Chern
number is ill defined for them.

In ZrTe the nodal ring around K and K' points acquires a small gap
(see also the inset in Fig.~\ref{fig:fp}(b)). Interestingly, three
pairs of WPs, slightly away from the $k_z = 0$ plane, form at each K
and K' point in ZrTe~\cite{xi_dai_ZrTe} and also in
MoP~\cite{MoP1}. WC (together with MoC and WN) remains a nodal line
metal (see inset Fig.~\ref{fig:fp}(d)). There exist two nodal rings
(one inside another) formed by two touching bands protected by the
horizontal mirror $\sigma_h$. For WN there is only a single such nodal
ring around each $K$ and $K'$. The stability of these topological
features has been checked with the HSE06 hybrid
functional~\cite{heyd2003hybrid,winkler2016triple}. Furthermore, the
band structure of MoP has been recently experimentally investigated
using angle-resolved photoemission spectroscopy (ARPES), confirming
experimentally the presence of TPs and WPs~\cite{MoP1}.

In Fig.~\ref{fig:fp}(g) a zoom-in of the $\Gamma$-A line in ZrTe is
shown. The Fermi level resides in between the $\Gamma_9$ and
$\Gamma_8$ bands at A. Upon turning on the SOC the no-SOC $\Lambda_3$
state splits into the singly degenerate $\Lambda_4+\Lambda_5$ states
and the doubly degenerate $\Lambda_6$ state. Another $\Lambda_6$ state
comes from the no-SOC $\Lambda_1$. The two $\Lambda_6$ states
hybridize and each of them crosses with the spin-split $\Lambda_{4,5}$
states creating 2 pairs of TPs: $(G_1,G_2)$ and $(G_3,G_4)$. Each TP
is protected by the $C_{3v}$ symmetry of the $\Gamma$-A line.
Fig.~\ref{fig:fp}(h-i) shows the dispersion in the (100) direction for
$k_z$ tuned to the position of $G_1$, $G_2$ respectively. A linear
band crossing superimposed with a quadratic band resembles a WP,
degenerate with a quadratic band, similar to the findings of
Ref.~\cite{winkler2016topological} for type-B TPTMs. Again, band
inversion is the mechanism leading to the formation of TPs.

In Fig.~\ref{fig:fp}(j-k) the surface states of ZrTe for the (010)
surface are shown~\cite{winkler2016triple, wu2017wanniertools}. The surface
potential is found to depend strongly on the termination choice: Zr
(Te)-termination is shown in Fig.~\ref{fig:fp}(j)
(Fig.~\ref{fig:fp}(k)). Since the $k_z = \pi$ plane is a quantum spin
Hall insulator plane, a Kramers doublet of surface states should
appear along the $\overline{\mathrm{A}}$-$\overline{\mathrm{R}}$ line
of the surface BZ. Consequently, there is a surface Dirac cone SS1
located at $\overline{\mathrm{A}}$ ($\overline{\mathrm{R}}$) for Zr
(Te) termination. The surface states forming the Dirac cone emerge
from the TPs $G_1$ and $G_2$. For $k_z$ values below the location of
$G_1$ and $G_2$, there exists another pair of surface states SS1'
emerging from the TPs. SS1', however, is not topologically protected.

The K' point of the bulk BZ is projected onto the
$\overline{\Gamma}$-$\overline{\mathrm{X}}$ line in
Fig.~\ref{fig:fp}(j-k) (compare to Fig.~\ref{fig:struct}(b)). A small
gap due to SOC can be visible in the projected bulk spectrum around
the projection of K point (shown with an arrow).  For ZrTe the $k_z=0$
mirror plane hosts a quantum spin Hall phase with the mirror Chern
numbers $\pm 1$, thus one can expect to see a Kramers pair of
topological surface state along the line
$\overline{\mathrm{X}}\leftarrow \overline{\Gamma}\rightarrow
-\overline{\mathrm{X}}$. This expectation is further supported by the
Berry curvature calculation in the $k_z = 0$ plane, see
Fig.~\ref{fig:mirror}. It reveals the accumulation of Berry curvature
in an area around the K (K') point that sums up to approximately $-1$
($1$)~\cite{winkler2016triple}. In accord with this topological
arguments one finds a quantum Hall like surface state SS2 crossing the
gap along $\overline{\Gamma}$-$\overline{\mathrm{X}}$ (its Kramers
partner is not shown, being at TR-symmetric part of the surface BZ.

Fig.~\ref{fig:fp}(l-m) show the (010)-surface Fermi surface revealing
double Fermi arcs between the two hole pockets containing the TPs,
corresponding to SS1 and SS1'. The state SS2 is not visible for this
choice of the Fermi level. For Te termination the Fermi arcs connect
the two hole pockets, while they do not touch them for the Zr
termination. In both cases, however, the surface states are protected
by TR and mirror symmetry on the $k_z = \pi$ plane, so they can not be
fully removed from the spectrum. ARPES investigations of the very
similar TPTM WC prove the existence of the surface Fermi arcs SS1 and
SS1'~\cite{WC_arpes}.

\subsection{  Cubic symmetry}
\begin{figure}
  \centerline{\includegraphics[width = \linewidth]{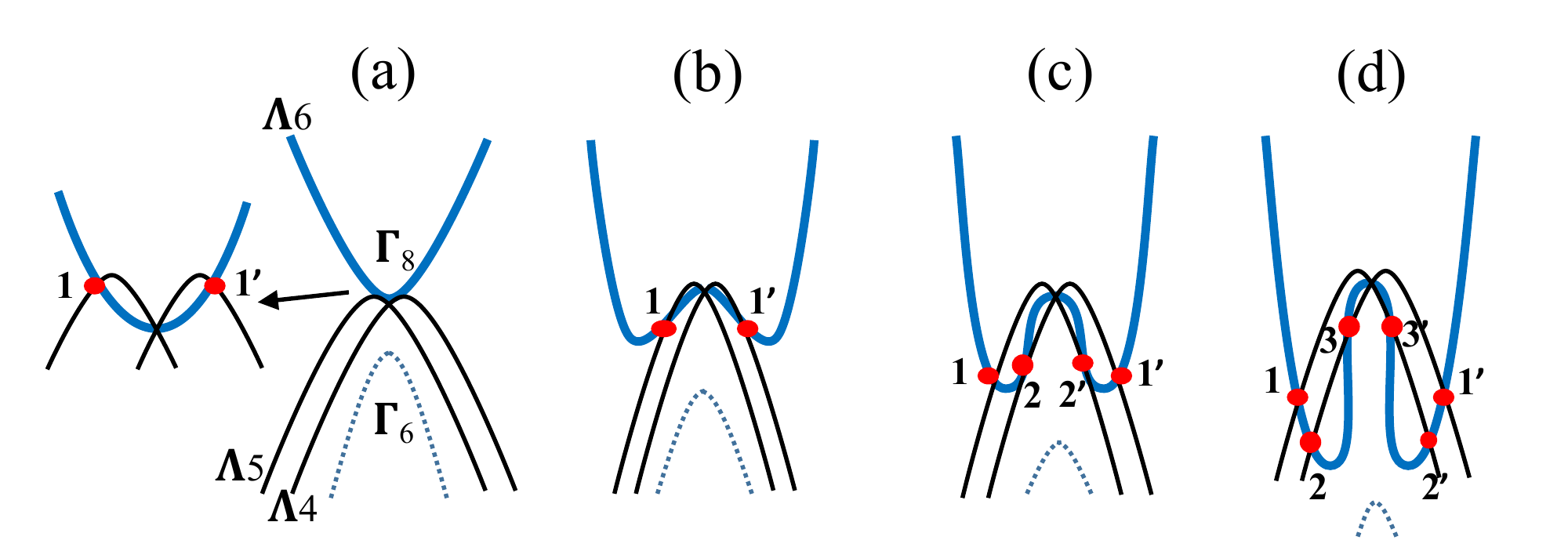}}
  \caption{\label{fig:cubic} Energy bands along L-$\Gamma$-L in cubic
    crystalls with band inversion. The thick blue line is doubly
    degenerate whereas black lines are non-degenerate. (a) HgTe-type
    energy bands. (b-d) Evolution of band structure in half-Heusler
    materials with increasing number of TPs. Illustration adapted with
    permission from Ref.~\cite{half_heusler}}
\end{figure}

One of the first account of TPs has been given in the context of HgTe
by Saad Zaheer et al. in Ref.~\cite{zaheer_hgte}. According to the
cubic symmetry there are eight equivalent directions with little group
$C_{3v}$ hosting possible TPs. Due to the lack of a horizontal mirror
these TPs are all of type-B. The situation is complicated by the fact
that the $\Lambda_{4,5}$ and $\Lambda_6$ states are degenerate at
$\Gamma$ forming the four-dimensional $\Gamma_8$ representation. This
leads generically to eight TPs near a $\Gamma_8$ crossing along the
eight equivalent $\Lambda$ directions, see
Fig.~\ref{fig:cubic}~(a). However, in the HgTe case the TPs are
extremely close to $\Gamma$ such that they are impossible to resolve
with available experimental techniques.

The same scenario takes place in half-Heusler compounds with band
inversion~\cite{yu_half_heusler}. Interestingly, in some compounds
like LuPtBi, LuAuPb and YPtBi the energy bands have such a strong
non-parabolicity that additional TPs appear~\cite{half_heusler}. The
evolution of the band structure with increasing number of TPs is
illustrated in Fig.~\ref{fig:cubic}~(b-d). Therefore these
half-Heusler compounds have three TPs in each equivalent 111
direction, 24 in total. Furthermore, these TPs are well separated from
the $\Gamma$ point with observable Fermi arcs on the easily cleavable
(111)
surface~\cite{half_heusler,half_heusler_surface1,half_heusler_surface2}. These
properties make the half-Heusler compounds ideally suited to
experimentally investigate the intricate topology and surface states
of type-B TPs.

\subsection{  CuPt-ordered \inassb{}}
\begin{figure}
  \centerline{\includegraphics[width = \linewidth]{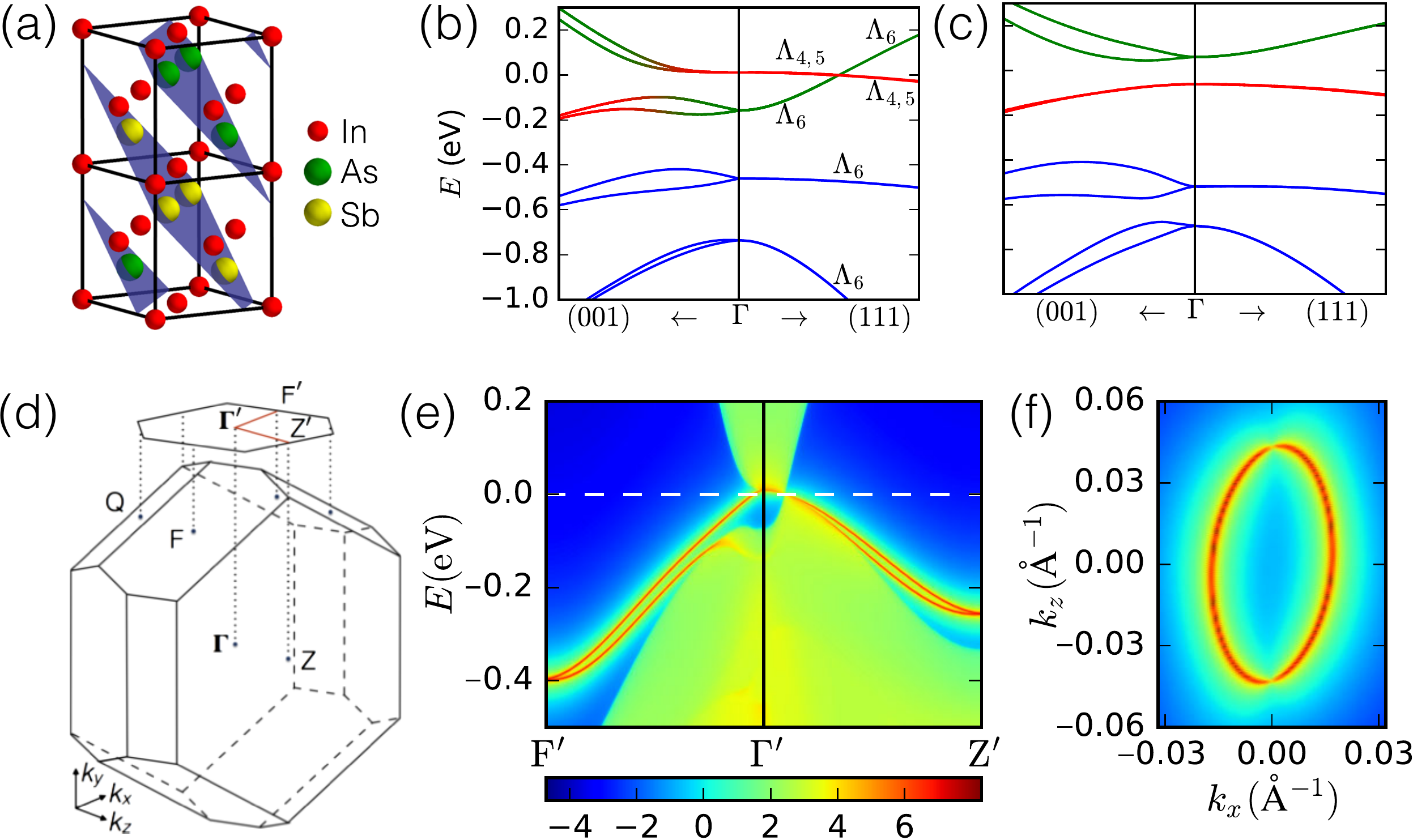}}
  \caption{\label{fig:inassb} (a) Crystal structure of CuPt-ordered
    \inassbh{}. (b) Band structure of CuPt-ordered \inassbh{} around
    $\Gamma$ (plotted up to $|{\bf k}|=0.1\,\mathrm{\AA}^{-1}$). (c)
    Band structure with 3\% compressive strain applied in the
    111-direction. (d) Brillouin zone and surface projection of
    CuPt-ordered \inassbh{}. (e) Surface density of states for the
    $(1\bar 1 0)$ surface. (f) Topological Fermi arcs on the
    $(1\bar 1 0)$ surface. Figure partially reused from
    Ref.~\cite{winkler2016topological}}
\end{figure}

Above we have seen that the TPs of the cubic HgTe are too close to
$\Gamma$ to enable a direct experimental observation. However, this
can be changed by applying moderate tensile strain in the 111
direction, such that the remaining point group is broken down from
$T_d$ to $C_{3v}$~\cite{zaheer_hgte}. In this case two pairs of TPs
survive. Unfortunately, it is experimentally difficult to apply 111
strain to HgTe. The ingredients to achieve well separated TPs in HgTe
are band inversion and breaking of cubic $T_d$ symmetry to
$C_{3v}$. It is no surprise that the same ideas can be applied in
other III-V semiconductors. The difficulty of applying 111 strain can
be overcome by changing the mechanism of symmetry breaking from strain
to alloy ordering. Fortunately, some ternary III-V compounds show
naturally a so-called CuPt-ordering~\cite{stringfellow_1991}, which is
ordering in alternating 111 planes, see
Fig.~\ref{fig:inassb}~(a). This type of ordering provides exactly the
desired symmetry breaking. Especially promising is the \inassb{}
system~\cite{winkler2016topological}. InAs and InSb are both small
band gap semiconductors and the band gap is further reduced in their
alloy~\cite{bowing_2008,bowing_2012,suchalkin_gap_2016}. CuPt-ordering
further reduces the band
gap~\cite{stringfellow_1989,stringfellow_1991,kurtz_1992} and
first-principles calculations show that perfectly CuPt-ordered
\inassbh{} alloy has an inverted band
order~\cite{zunger_inversion_1991,winkler2016topological}. While
perfect atomic CuPt-ordering in \inassbh{} is difficult to obtain the
same effect is achieved by growing layered structures of \inassb{}
with alternating alloy composition $x$. With this approach the band
gap can be tuned from normal to inverted band
ordering~\cite{suchalkin_ordering_2015,InAsSb_experiment}.

The band structure of CuPt-ordered \inassbh{} obtained from
first-principles simulations is shown in
Fig.~\ref{fig:inassb}(b)~\cite{winkler2016topological}.  A pair of TPs
is formed at the crossings of the $\Lambda_{4,5}$ with the $\Lambda_6$
band.  The $\Lambda_{4,5}$ bands have a very small linear in $k$,
splitting, which is only about 2 meV at the momentum of the TPs. The
pair of TPs is thus very close together and will appear like a single
Dirac point in low resolution experiments.

Another very interesting aspect of CuPt-ordered \inassb{} is that the
symmetry breaking to $C_{3v}$ simultaneously allows for a Rashba-type
spin-orbit splitting in the conduction band, provided the band order
is normal. Either by imperfect CuPt-ordering, or by 111 strain, the
normal band order can be restored and one obtains a giant-Rashba
material, see Fig.~\ref{fig:inassb}~(c). Large Rashba splitting is a
crucial ingredient for topological quantum computers relying on
non-abelian
Majorana~\cite{ivanov,reed_green,kitaev,top_comp1,top_comp2}
quasiparticles realized in semiconductor
nanowires~\cite{Lutchyn,Oreg}. Since InAs and InSb are the preferred
materials for the experimental realizations of Majorana
quasiparticles~\cite{das_majorana_2012,xu_majorana_2012,mourik_majorana_2012,CM_Majorana2016,Zhang2016},
CuPt-ordered \inassb{} seems to be a very promising material for this
application~\cite{winkler2016topological}.

Fig.~\ref{fig:inassb}~(e) and (f) shows the surface density of states
of the $(1\bar 1 0)$ surface~\cite{winkler2016topological}. In this
case the Fermi arcs resulting from the TPs look very similar to those
from the typical Dirac
semimetals~\cite{wang_dirac_2012,wang_cd3as2,Xu_dirac_arpes} due to
the small separation of the TPs.

\section{Topological properties of Triple-Point Fermions}

In this section we first introduce simplified models of TP materials
and use them to discuss the topological properties of TPs.

\subsection{  Microscopic models}

Before we investigate the topology of TPs we introduce a few simple
$\kp$-models for type-A and type-B triple-points. The type-A
$\kp$-models are up to a change of basis identical to the ones in
Ref.~\cite{winkler2016triple} and the given parameters correspond to
the material realization of ZrTe. The type-B $\kp$-models are also up
to a basis transformation identical to the ones in
Ref.~\cite{winkler2016topological} and the parameters correspond to
the CuPt-ordered \inassbh{} case.

\subsubsection{Type-A: ZrTe}
For the WC-class of type-A TPTMs we construct a model around the A
point which captures the band inversion and describes the TPs. We
include the $\Gamma_9$, $\Gamma_8$ and $\Gamma_7$ states (see
Fig.~\ref{fig:fp}~(g) and Tab.~65 of Ref.~\cite{koster1963properties})
with energies close to the Fermi level. Along the $z$-axis the
$\Gamma_9$ representations splits into the $\Lambda_{4,5}$
representations and both $\Gamma_{7,8}$ become $\Lambda_6$. The little
group of A is $D_{3h}$ plus TR symmetry. For the derivation of the
$\kp$ models we need to identify the correct representations of the
symmetry operations. The Hamiltonian is then constructed such that it
commutes with all symmetries $S$
\begin{equation}
  H(S(\bk)) = R_S H(\bk) R_S^\dag,
  \label{eq:symcon}
\end{equation}
with $R_S$ being the representation of S in the basis of $H$.
Since all representations are two dimensional the symmetry
representation are the direct sum of two dimensional representations
$R(\Gamma_9) \oplus R(\Gamma_7) \oplus R(\Gamma_8)$,
\begin{equation}
  \begin{aligned}
    C_3 &= \begin{pmatrix} -\frac{1}{2} & -\frac{\sqrt{3}}{2} \\
      \frac{\sqrt{3}}{2} & -\frac{1}{2} \\ && 1 \end{pmatrix},\
    R_{C_3} = \mathrm{diag}(-1, -1, e^{-i\pi/3}, e^{i\pi/3}, e^{-i\pi/3}, e^{i\pi/3}),\\
    \sigma_v &= \mathrm{diag}(-1,\phantom{-}1,\phantom{-}1),\
    R_{\sigma_v} = \phantom{-}i\, \mathrm{diag}(\phantom{-}1 ,\phantom{-}1, \phantom{-}1) \otimes \tau_x ,\\
    \sigma_h &= \mathrm{diag}(\phantom{-}1,\phantom{-}1,-1),\
    R_{\sigma_h} = \phantom{-}i\, \mathrm{diag}(\phantom{-}1 ,\phantom{-}1, -1) \otimes \tau_z,\\
    \mathrm{TR} &= \mathrm{diag}(-1,-1,-1),\ R_\mathrm{TR} = -i\,
    \mathrm{diag}(\phantom{-}1 ,\phantom{-}1, \phantom{-}1) \otimes
    \tau_y ,
  \end{aligned}
\end{equation}
with $\tau_x$, $\tau_y$ and $\tau_z$ being the Pauli matrices.

Considering the constraint Eq.~\eqref{eq:symcon} for all symmetries
above, one obtains the following Hamiltonian
\begin{equation}
  {\scriptsize
    H_{\kp}^{\mathrm{A}} = \begin{pmatrix}
      \epsilon_{1} & A k_{z} & E \left(k_{x} - i k_{y}\right) & 0 & 0 & D \left(- i k_{x} + k_{y}\right)\\A k_{z} & \epsilon_{1} & 0 & E \left(- k_{x} - i k_{y}\right) & D \left(i k_{x} + k_{y}\right) & 0\\E \left(k_{x} + i k_{y}\right) & 0 & \epsilon_{2} & 0 & - i B k_{z} & C \left(- i k_{x} - k_{y}\right)\\0 & E \left(- k_{x} + i k_{y}\right) & 0 & \epsilon_{2} & C \left(i k_{x} - k_{y}\right) & - i B k_{z}\\0 & D \left(- i k_{x} + k_{y}\right) & i B k_{z} & C \left(- i k_{x} - k_{y}\right) & \epsilon_{3} & 0\\D \left(i k_{x} + k_{y}\right) & 0 & C \left(i k_{x} - k_{y}\right) & i B k_{z} & 0 & \epsilon_{3}
    \end{pmatrix},}
  \label{eq:kpa}
\end{equation}
using the definition
\mbox{$\epsilon_i({\bf k}) = E_i + F_i (k_x^2+k_y^2) + G_i k_z^2$} and
$\bk$ relative to the A point. Via fitting to the ZrTe band structure
we obtain the following parameters for Eq.~\eqref{eq:kpa}:
$E_1 = -0.0391$ eV, $E_2 = 1.3709$ eV, $E_3 = 0.0391$ eV, $F_1 = 3.75$
eV\AA$^2$, $F_2 = -0.5$ eV\AA$^2$, $F_3 = 4.25$ eV\AA$^2$, $G_1=2.2$
eV\AA$^2$, $G_2 = -12.64$ eV\AA$^2$, $G_3 = 1.5$ eV\AA$^2$, $A=0.17$
eV\AA, $B=0.24$ eV\AA, $C=2.9$ eV\AA, $D=0.054$ eV\AA{} and $E=2.76$
eV\AA.

If the influence of the $\Gamma_7$ state is removed one obtains a
$4\times 4$ model
\begin{equation}
  {
    H_{\kp}^{\mathrm{A,}4\times 4} = \begin{pmatrix}
      \epsilon_{1} & A k_{z} & 0 & D \left(- i k_{x} + k_{y}\right)\\A k_{z} & \epsilon_{1} & D \left(i k_{x} + k_{y}\right) & 0\\0 & D \left(- i k_{x} + k_{y}\right) & \epsilon_{3} & 0\\D \left(i k_{x} + k_{y}\right) & 0 & 0 & \epsilon_{3}
    \end{pmatrix}.  }
  \label{eq:kpaa}
\end{equation}
To simulate the interaction of the two $\Lambda_6$ bands we add a
fourth order term to
$\epsilon_3({\bf k}) = E_3 + F_3 (k_x^2+k_y^2) + G_3 k_z^2 + H_3
k_z^4$. We found the following parameters via fitting to the band
structure of ZrTe: $E_1 = -0.0391$ eV, $E_3 = 0.0391$ eV, $F_1 = 4.5$
eV\AA$^2$, $F_3 = -7.3$ eV\AA$^2$, $G_1=2.2$ eV\AA$^2$, $G_3 = 3.2$
eV\AA$^2$, $H_1 = 0$ eV\AA$^4$, $H_3 = -20$ eV\AA$^4$, $A=0.17$
eV\AA{} and $D=0.487$ eV\AA. Note that the parameter A is the only one
that breaks inversion symmetry in the above model. Setting $A=0$ one
obtains a $\kp$ description of a Dirac semimetal.

A four band Hamiltonian in the vicinity of a pair of TPs, where $\bk$
is understood relative to the midpoint of the two TPs is given below
\begin{equation}
  {
    H_{\kp}^{\mathrm{TP}_\mathrm{A}} = \begin{pmatrix}
      A_{1} k_{z} & A_{2} k_{z} + E_{0} & i D \left(k_{x} - i
        k_{y}\right) & i C \left(k_{x} + i k_{y}\right)\\
      A_{2} k_{z} + E_{0} & A_{1} k_{z} & i C \left(k_{x} - i k_{y}\right) & i D
      \left(- k_{x} - i k_{y}\right)\\
      -i D \left(k_{x} + i k_{y}\right) & -i C \left(k_{x} + i
        k_{y}\right) & B k_{z} & 0\\
      -i C \left(k_{x} - i k_{y}\right) &
      -i D \left(- k_{x} + i k_{y}\right) & 0 & B k_{z}
    \end{pmatrix}.}
  \label{eq:kpac}
\end{equation}
Instead of $\sigma_h$ and TR only their product
$\theta \circ \sigma_h$ needs to be taken into account at a general
$\bk$-point on the $C_{3v}$-symmetric $z$-axis. In the following we
used $E_0 = 30\,\mathrm{meV}$, $A_1 = 1.4\,\mathrm{eV\AA}$,
$A_2=0\,\mathrm{eV\AA}$, $B=-1.0\,eV\mathrm{\AA}$,
$C = 1.082\,eV\mathrm{\AA}$ and $D=0\,\mathrm{eV\AA}$. We use above
$\kp$ model for the illustrations of the type-A TPTM in
Fig.~\ref{fig:nodal_lines}.

The WC-class of TPTMs also has interesting topological properties
around the K-points. A good $\kp$ description of the topology and
bands around K (or K') requires at least 8 states. The little group of
the K points is $C_{3h}$ and the $\Gamma_7$, $\Gamma_{12}$,
$\Gamma_{11}$, $\Gamma_9$, $\Gamma_{12}$, $\Gamma_{10}$, $\Gamma_8$
and $\Gamma_7$ states (see Tab.~57 of
Ref.~\cite{koster1963properties}) are determined to be relevant for
constructing a $\kp$-description. We use the following symmetry
representations
\begin{equation}
  \begin{aligned}
    C_3 &= \begin{pmatrix} -\frac{1}{2} & -\frac{\sqrt{3}}{2} \\  \frac{\sqrt{3}}{2} & -\frac{1}{2}  \\ && 1 \end{pmatrix}, R_{C_3} = \mathrm{diag}(e^{i\frac{\pi}{3}},-1,-1,e^{i\frac{\pi}{3}},-1,e^{-i\frac{\pi}{3}},e^{-i\frac{\pi}{3}},e^{i\frac{\pi}{3}}) ,\\
    \sigma_h &= \mathrm{diag}\{\phantom{-}1,\phantom{-}1,-1\},
    R_{\sigma_h} = \mathrm{diag}( i,-i,i,-i,-i,i,-i,i) .
  \end{aligned}
\end{equation}

Considering the symmetries given above, the lowest order Hamiltonian
around K is given by
\begin{equation}
  {
    H_{\kp}^\mathrm{K} = \begin{pmatrix}
      \epsilon_1 & 0 & B_1\,k^+ & A_1\,k_z & 0 & B_3\,k^- & 0 & 0 \\
      0 & \epsilon_2 & -A'_1\,k_z & B_2\,k^- & 0 & 0 & B_4\,k^+ & 0 \\
      B_1^* \, k^- & -A^{\prime *}_1\,k_z & \epsilon_3 & 0 & A_2\,k_z & B_9\,k^+ & 0 & B_5\,k^- \\
      A_1^*\,k_z & B_2^*\,k^+ & 0 & \epsilon_4 & -B'_9\,k^+ & 0 & B_6\,k^- & A_4\,k_z \\
      0 & 0 & A_2^* \,k_z & -B^{\prime *}_9\,k^- & \epsilon_5 & 0 & B_7\,k^+ & 0 \\
      B_3^*\,k^+ & 0 & B_9^*\,k^+ & 0 & 0 & \epsilon_6 & A_3\,k_z & B_8\,k^+ \\
      0 & B_4^*\,k^- & 0 & B_6^*\,k^+ & B_7^*\,k^- & A_3^*\,k_z & \epsilon_7 & 0 \\
      0 & 0 & B_5^*\,k^+ & A_4^*\,k_z & 0 & B_8^*\,k^- & 0 & \epsilon_8
    \end{pmatrix},}
  \label{eq:kpk}
\end{equation}
using $\epsilon_i$ defined as in Eq.~\eqref{eq:kpa},
$k^\pm = k_x \pm i k_y$ and $\bk$ relative to K. Since K is not a
time-reversal invariant momentum bands do not form doubly degenerate
Kramers pairs at this point.  For the $\kp$ model around the K point
we obtain the following parameters (in units of eV and \AA) via
fitting to the ZrTe band structure: $E_1 = -0.0979$, $E_2 = -0.0671$,
$E_3 = 0.6538$, $E_4 = 0.8393$, $E_5 = 1.0661$, $E_6 = 1.1351$,
$E_7 = 1.2145$, $E_8 = 1.2774$, $F_1 = F_2 = 3.6$, $F_3=F_4 = -2.0$,
$F_5=F_6= 6.0$, $F_7=F_8=1.5$, $G_1 = G_2 = 3.6$, $G_3 = G_4 = -0.2$,
$G_5 = G_6=2.0$, $G_7=G_8=-3.0$, $A_1 = A'_1= 4.0$, $A_2 = 0.2$,
$A_3 = 0$, $A_4 = 0$, $B_1 = 0.2-i 0.1$, $B_2 = 0.02-i 0.01$,
$B_3 = 0.2$, $B_4 = -0.2$, $B_5 = -1.0 + i 4.0$, $B_6 = -4.0+i 1.0$,
$B_7 = 3.0 + i 0.5$, $B_8 = -0.5 + i 3.0$ and $B_9 = 1.5$. The Weyl
points reported in Ref.~\cite{xi_dai_ZrTe} are also described by this
$\kp$ model.

\subsubsection{Type-B: CuPt-ordered \inassbh{}}
The type-B scenario differs only by the absence of the horizontal
mirror $\sigma_h$. For completenes we give the corresponding
representations again, albeit $\sigma_v$ is now a mirror in the
$xz$-plane instead of the $xy$-plane as before. The representations
below correspond to a $\Lambda_{4,5}\oplus\Lambda_{6}$-basis
\begin{equation}
  \begin{aligned}
    C_3 &= \begin{pmatrix} -\frac{1}{2} & -\frac{\sqrt{3}}{2} \\
      \frac{\sqrt{3}}{2} & -\frac{1}{2} \\ && 1 \end{pmatrix},\
    R_{C_3} = \mathrm{diag}(-1, -1, e^{-i\pi/3}, e^{i\pi/3}),\\
    \sigma_v &= \mathrm{diag}(\phantom{-}1,-1,\phantom{-}1),\
    R_{\sigma_v} = \phantom{-}i\, \mathrm{diag}(\phantom{-}1 ,\phantom{-}1) \otimes \tau_y ,\\
    \mathrm{TR} &= \mathrm{diag}(-1,-1,-1),\ R_\mathrm{TR} = -i\,
    \mathrm{diag}(\phantom{-}1 ,\phantom{-}1, \phantom{-}1) \otimes
    \tau_y ,
  \end{aligned}
\end{equation}
Around $\Gamma$ the following model is obtained
\begin{equation}
  {
    H_{\kp}^\mathrm{B} = \begin{pmatrix}
      \epsilon_{1} & - i A k_{z} & i B_1 \left(k_{x} - i k_{y}\right)
      & i B_2 \left(k_{x} + i k_{y}\right)\\
      i A k_{z} & \epsilon_{1} & i B_2 \left(k_{x} - i k_{y}\right) &
      i B_1 \left(- k_{x} - i k_{y}\right)\\
      -i B_1 \left(k_{x} + i k_{y}\right) & -i B_2 \left(k_{x} + i
        k_{y}\right) & \epsilon_{2} & i C \left(k_{x} - i
        k_{y}\right)\\
      -i B_2 \left(k_{x} - i k_{y}\right) & -i B_1 \left(- k_{x} + i
        k_{y}\right) & -i C \left(k_{x} + i k_{y}\right) & \epsilon_{2}
    \end{pmatrix}, }
  \label{eq:kpb}
\end{equation}
with the following set of parameters for CuPt-ordered \inassbh{}:
$E_1 = 0.1696$ eV, $E_2 = 0$ eV, $F_1=-5.86$ eV\AA$^2$, $F_2=14.3$
eV\AA$^2$, $G_1=-3.91$ eV\AA$^2$, $G_2=54.2$ eV\AA$^2$, $A=0.02$
eV\AA, $B_1=0$$, $$B_2=1.48$ eV\AA{} and $C=1.26$ eV\AA.

Omitting the TR symmetry, a Hamiltonian of a pair of TPs, with $\bk$
relative to the midpoint of the two TPs, is obtained
\begin{equation}
  {
    H_{\kp}^{\mathrm{TP}_\mathrm{B}} = \begin{pmatrix}
      A_{1} k_{z} & - i \left(A_{2} k_{z} + E_{0}\right) & F \left(k_{x}
        - i k_{y}\right) & D \left(k_{x} + i k_{y}\right)\\
      i \left(A_{2}
        k_{z} + E_{0}\right) & A_{1} k_{z} & D \left(- k_{x} + i
        k_{y}\right) & F \left(k_{x} + i k_{y}\right)\\
      F^* \left(k_{x} +
        i k_{y}\right)  & D^* \left(- k_{x} - i k_{y}\right) & B k_{z} & C
      \left(i k_{x} + k_{y}\right)\\
      D^* \left(k_{x} - i k_{y}\right) & F^*
      \left(k_{x} - i k_{y}\right) & C \left(- i k_{x} + k_{y}\right) & B k_{z}
    \end{pmatrix}.}
  \label{eq:kpbc}
\end{equation}
In contrast to Eq.~\eqref{eq:kpac} the parameters $D$ and $F$ can now
take complex values. We use $E_0=0.88$ meV, $A_1=-0.42$ eV\AA, $A_2=0$
eV\AA, $B=4.22$ eV\AA, $C=0.78$ eV\AA, $D=0.755-0.445i$ and
$F=-1.69+0.755i$. We use above $\kp$ model for the illustrations of
the type-B TPTM in Fig.~\ref{fig:nodal_lines}.

\subsection{  Wilson loop characterization for pairs of triple-points}
\label{sec:wilson}

The tracking of Wilson loop eigenvalues~\cite{yu_wilson}, or
equivalently the so-called hybrid Wannier
centers~\cite{resta_hybrid_2001}, under an adiabatic modification of a
Hamiltonian has been proven as an invaluable tool for classifying
topological phases~\cite{Fu_time_2006, alexey_invariants_2011,  alexey_weyl, gresch}. Here a $\mathbb Z_2$ topological
classification based on the Wilson loop is developed for pairs of TPs.

The Wilson loop can be defined on any path in $k$-space connecting two
points ${\bf k_1}$ and ${\bf k_2}$ with the property
${\bf k_1} = {\bf k_2 } + {\bf G}$, where ${\bf G}$ is a reciprocal
lattice vector. The Wilson loop is defined as the path ordered
product~\cite{yu_wilson}
\begin{equation}
  \mathcal{W}_{\bf k_1 k_2} = P_{\bf k_1} \left( \prod_{j=1,2,\dots} P_{\bf k'_j} \right) P_{\bf k_2},
  \label{eq:wilson}
\end{equation}
with
\mbox{$P_{\bf k} = \sum_{n\in \mathrm{occ.}} |u_n({\bf k})\rangle
  \langle u_n({\bf k}) |$} the projector on the occupied subspace of a
Hamiltonian. The Wilson loop is inherently gauge invariant due to the
gauge invariance of the projector $P_{\bf k}$. The Berry phase
associated with the loop is given by the determinant of the Wilson
operator \mbox{$\det(\mathcal{W}) = \exp(i\phi_B)$}. If the
Hamiltonian has a symmetry $R$, it can be shown
that~\cite{andrei_point_group}
\begin{equation}
  \tilde R \mathcal{W}_{\bf k_1 k_2} \tilde R^{-1} = \mathcal{W}_{R{\bf k_1}R{\bf k_2}},
  \label{eq:wsym}
\end{equation}
with $R$ acting in reciprocal space and $\tilde R$ in occupied band
space. The symmetry expectation value of Wilson loop eigenstates
$|v_i \rangle$ is calculated as
$\langle v_i | \tilde R | v_i \rangle$.

For Weyl~\cite{alexey_weyl} and Dirac~\cite{gresch} semimetals it is
known that the Wilson loop spectrum on a sphere enclosing the
semimetallic point gives the topological classification of the
crossing. Also in our case with TPs a similar kind of topological
classification is possible. Here the classification is demonstrated
with a tight-binding model of ZrTe obtained in
Ref.~\cite{winkler2016triple}.
\begin{figure}
  \includegraphics[width = \linewidth]{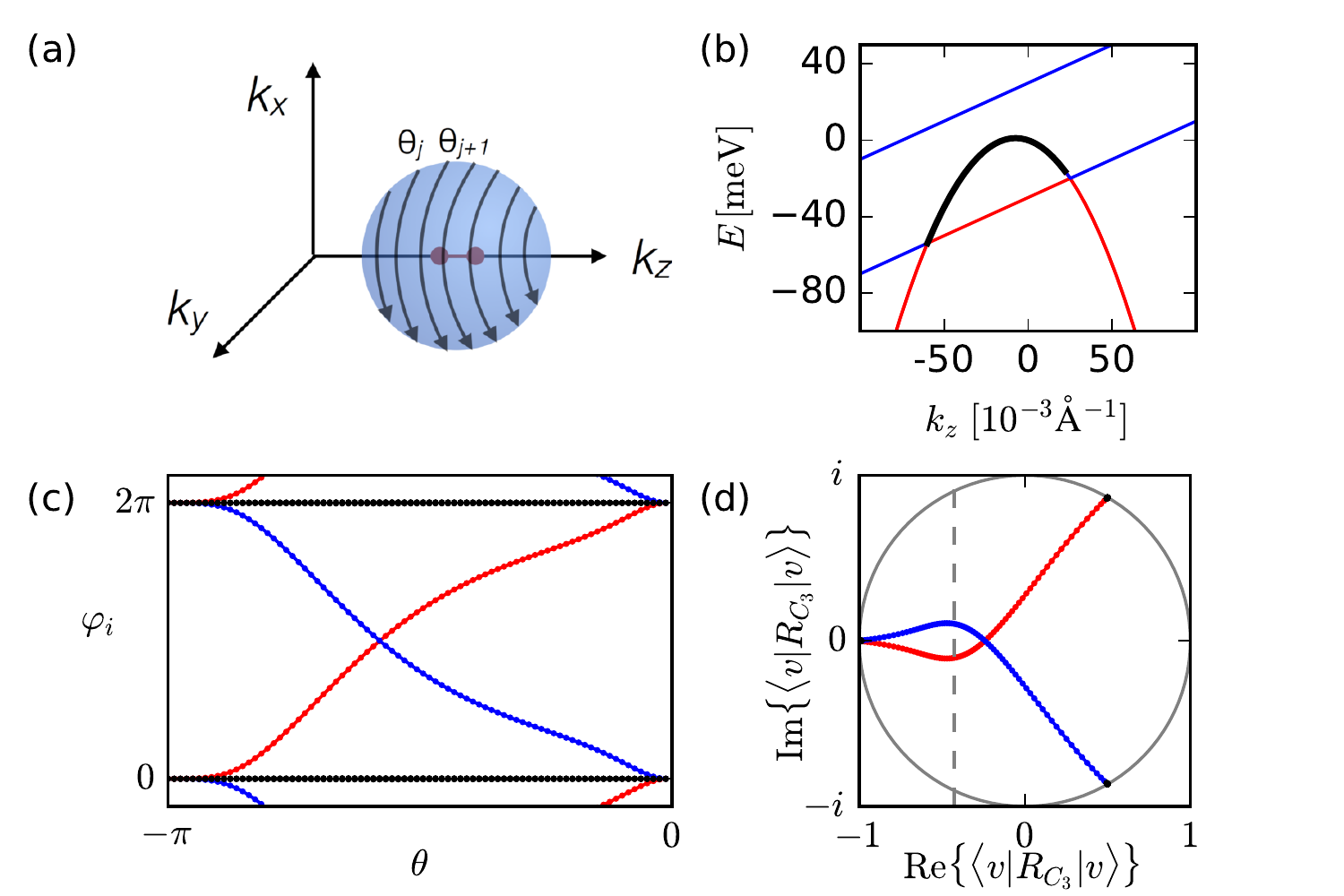}
  \caption{ (a) Pair of TPs (red points)
    connected by a nodal line (red line) enclosed by a sphere. The
    arrows indicate the individual Wilson loops winding around the
    sphere. (b) Example of a trivial pair of TPs. (c) The Wilson loop
    spectrum on a sphere enclosing a pair of TPs. Two Wilson loop
    eigenvalues feature gapless flow (colored in red and blue). (d)
    $C_3$ symmetry expectation value of individual Wilson
    lines. Figure reused from Ref.~\cite{winkler2016triple}.}
  \label{fig:wilson}
\end{figure}

In Fig.~\ref{fig:wilson}(a) we show a spherical surface on which the
Wilson loop spectrum is to be evaluated. The sphere is chosen such
that the Hamiltonian is gapped everywhere on the surface, the symmetry
axis containing the TPs goes through the center of the sphere and both
TPs are enclosed by the sphere. The latter point is important, since
there is always at least one nodal line connecting two TPs, therefore
including only one TP would not fulfill the requirement that the
Hamiltonian is gapped on the sphere. Note that the Wilson lines are
oriented such that the symmetry axis goes through their center. In
Fig.~\ref{fig:wilson}(c) we plot the phases $\phi_i$ of the individual
Wilson loop eigenvalues as a function of the azimuthal angle
$\theta$. The tight-binding model has 8 occupied states, therefore, we
obtain 8 Wilson loop eigenvalue phases $\phi_i$. 6 $\phi_i$ (marked in
black) are trivial and stay very close to 0 ($2\pi$), but two (marked
in red and blue) seem to cross. Note that the $\sigma_v$ symmetry
constrains the $\phi_i$ such that the Wilson loop spectrum is mirror
symmetric $\phi_i =
-\phi_j$~\cite{alexandradinata_inversion_2014}. Since the Hamiltonian
is gapped on the surface, and the Wilson loop is gauge invariant, the
individual $\phi_i$ change smoothly with $\theta$. Therefore, the
connectivity of the $\phi_i$ can be determined as long as they are not
degenerate. To obtain the connectivity across the degeneracy point
between the red and blue Wilson eigenvalues we calculate the $C_3$
symmetry expectation values of the corresponding states in
Fig.~\ref{fig:wilson}(d). The grey dashed line in
Fig.~\ref{fig:wilson}(d) indicates the position of the crossing of the
blue and red line in Fig.~\ref{fig:wilson}(c). Note that the crossing
of red and blue lines in Fig.~\ref{fig:wilson}(d) is accidental and we
found that it can be avoided via choosing a cigar-shape, rather than a
sphere. However, the $C_3$ symmetry expectation value is nondegenerate
at the crossing Fig.~\ref{fig:wilson}(c) and we can use
Fig.~\ref{fig:wilson}(d) to unambiguously determine the connectivity
for all $\theta$. Therefore, the red and blue lines in the Wilson loop
spectrum clearly indicate two hidden Berry curvature fluxes, one
inward and one outward, through the sphere. The fluxes can be
separated in the Wilson loop eigenbasis, corresponding to individual
Chern numbers~\cite{alexey_smooth} of $\pm 1$. The difference of the
two individual Chern numbers divided by two constitutes a
$\mathbb Z_2$ topological invariant for TPs.

At the polar regions $\theta \approx 0$ or $\theta \approx -\pi$ the
Wilson loop commutes with the $C_3$ symmetry due to
Eq.~\eqref{eq:wsym}. In this case the $C_3$ expectation value in
Fig.~\ref{fig:wilson}(d) is one of the possible $C_3$ eigenvalues
$\{-1,\exp(i\pi/3),\exp(-i\pi/3)\}$, which are the starting and ending
points of the lines in Fig.~\ref{fig:wilson}(d). Note that the 6
trivial $\phi_i$ (black dots) are almost fixed to the $C_3$
eigenvalues, whereas the two nontrivial $\phi_i$ (red/blue dots)
change the $C_3$ eigenvalue from $\{-1,\, -1\}$ to
$\{\exp(i\pi/3),\, exp(-i\pi/3)\}$. Responsible for this behaviour are
the two valence bands having the rotational eigenvalues $-1,\, -1$ for
$k_z$ to the left of the two TPs and $\exp(i\pi/3),\, exp(-i\pi/3)$
for $k_z$ to the right of the two TPs. Therefore, the planes above
$G_1,\, G_2$ are topologically distinct from the planes below,
consequently uncovering the existence of crossing points realized as
the two TPs here.

In Fig.~\ref{fig:wilson}(b) we give an example of a topologically
trivial pair of TPs. In this case the $C_3$ eigenvalues of the valence
bands are the same to the left and to the right of the two TPs and
hence the Wilson loop spectrum is in general gapped with an even
$\mathbb Z_2$ invariant.

\subsection{  Lifshiftz transitions}

\begin{figure}
  \includegraphics[width = \linewidth]{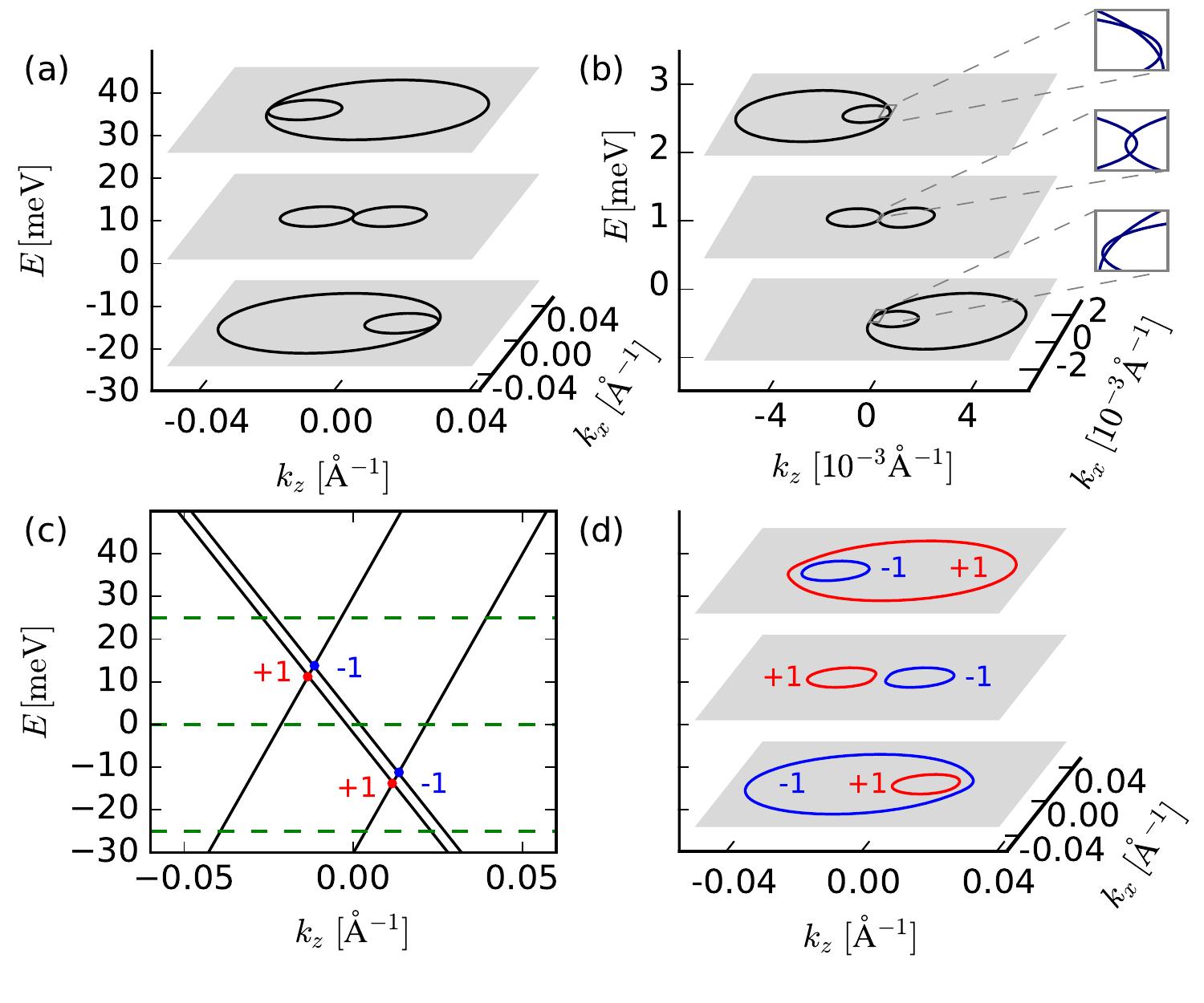}
  \caption{(a)((b)) Fermi surfaces for type-A
    (type-B) triple-point topological metals at three different energy
    cuts: below, between and above the two triple-points. The three
    small insets in panel (b) show that for the type-B scenario there
    are several distinct touching points between the Fermi
    pockets. (c) Band structure around the triple-points for a small
    Zeeman field parallel to the $C_3$ axis. (d) Fermi surface of
    type-A triple-point topological metal with a small Zeeman
    field. In panel (c) and (d) the Chern numbers of WPs and Fermi
    surfaces are marked in red ($+1$) and blue ($-1$). Figure reused
    from Ref.~\cite{winkler2016triple}.}
  \label{fig:fermi_simple}
\end{figure}

Using the models of Eq.~\eqref{eq:kpac} and \eqref{eq:kpbc} we can
analyze the Lifshitz transitions of Fermi surfaces in the TPTMs. The
nodal lines of Fig.~\ref{fig:nodal_lines}(a-b) guarantee that several
Fermi surfaces touch within a finite energy window in between two
TPs. Fig.~\ref{fig:fermi_simple}(a) illustrates the fixed $k_y=0$ cuts
of the Fermi surface for the Fermi level $E_{\rm F}$ placed above,
below and in between the two TPs, representing three topologically
distinct Fermi surfaces.  At each of the two TPs a topological
Lifshitz transition takes place: one of the Fermi pockets shrinks to a
point reopening either inside or outside another Fermi pocket. When
the $E_{\rm F}$ is placed in between the two TPs there appears a
topologically protected touching point between electron and hole
pockets, similar to the type-II WP scenario~\cite{alexey_weyl}.

The Lifshitz transitions occurring in type-B TPTMs are illustrated in
Fig.~\ref{fig:fermi_simple}(b). The difference to the type-A
transitions is that a single touching point between the Fermi pockets
(the point of quadratic band touching) now splits into four points (or
two linear band touchings on each mirror plane) due to the breaking of
$\sigma_h$ (see insets of Fig.~\ref{fig:fermi_simple}(b)). Intricate
spin textures with changing winding numbers, were predicted for a
(111)-strained HgTe in Ref.~\cite{zaheer_hgte}, which according to the
classification is a type-B TPTM. Nontrivial windings in the
spin-texture are also found for type-A TPTMs~\cite{xi_dai_triple}.

Since the distinct Fermi pockets touch in TPTMs for a range of
energies, the topological charge of individual pockets is
undefined. However, as mentioned above, this degeneracy is lifted by
breaking $\sigma_v$ by, for example, applying a small Zeeman field in
the $z$ direction or photoirradiation effects~\cite{ezawa}. In this
case each of the TPs splits into two WPs with opposite Chern numbers
as illustrated in Fig.~\ref{fig:fermi_simple}(c). The touching Fermi
pockets now separate, and well-defined Chern numbers can be assigned
to each of them, see Fig.~\ref{fig:fermi_simple}(d). The Chern number
of a pocket is equal to the summed up chiralities of WPs enclosed
within it.

\subsection{  Surface states}
Here we use the previously defined $\kp$ models to gain insights into
the first principles surface states shown in
Sec.~\ref{sec:materials}. The $\kp$ surface state calculations are
faciliated by discretizing the momentum perpendicular to the surface,
and thus generating a 1D tight-binding model with the parallel momenta
as auxilliary parameters~\cite{kp_discretisation}. The SDOS is then
calculated using the iterative Green's function
method~\cite{sancho1984quick,sancho1985highly}. For the $\kp$ models
given in Eq.~\eqref{eq:kpa} and \eqref{eq:kpaa} we use 1~\AA{} as the
discretization length and 2~\AA{} for Eq.~\eqref{eq:kpk}.
\begin{figure}
  \includegraphics[width = \linewidth]{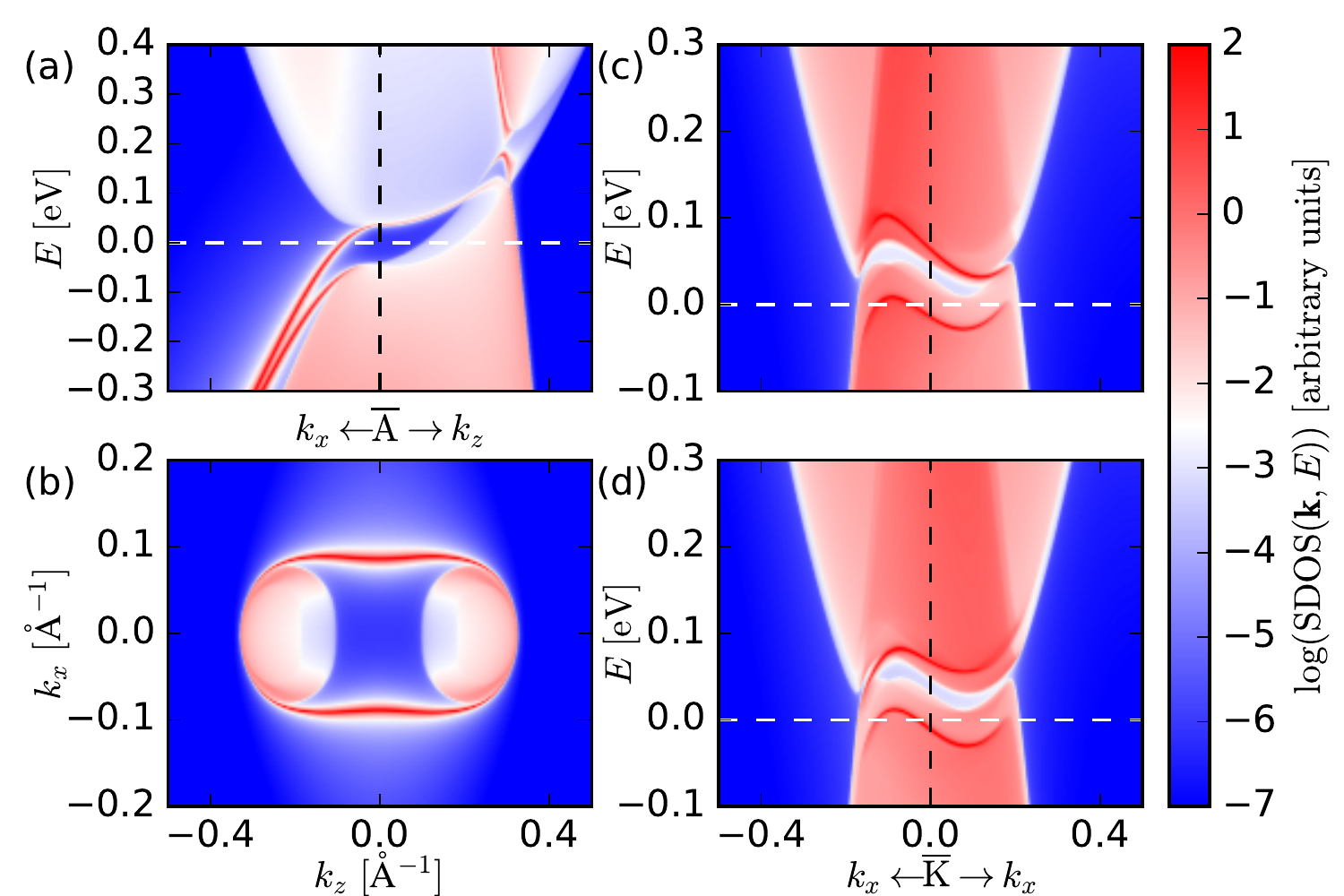}
  \caption{SDOS of the (010)-surface for the $\kp$ models given in
    Eq.~\eqref{eq:kpa} and \eqref{eq:kpk}. (a) and (b) show the SDOS
    and Fermi surface around the $\overline{\mathrm{A}}$ point. (c)
    ((d)) show the SDOS around the $\overline{\mathrm{K}}$ point for
    the top (bottom) surface. Figure reused from
    Ref.~\cite{winkler2016triple}.}
  \label{fig:kp_sdos}
\end{figure}

We use the $\kp$ models with parameters such that they fit the band
structure of ZrTe. The model around A, given in Eq.~\eqref{eq:kpa}, is
characterized by the mirror Chern numbers $C_{m=\pm i} = \mp 1$ in the
$k_z=\pi$ plane. Therefore, a topological-insulator-like surface state
is expected on a surface orthogonal to the $\sigma_h$ mirror plane. In
Fig.~\ref{fig:kp_sdos}(a) we show the SDOS on a surface orthogonal to
$y$, corresponding to the (010) surface in the WC structure. On the
$k_x$ axis the upper topologically nontrivial surface state emerges
from the conduction bands and connects to the valence bands. There is
another trivial surface state with opposite mirror eigenvalue
below. If we compare this to the first-principles surface states shown
in Fig.~\ref{fig:fp}(k) then these two surface states will form a
Dirac cone at $\overline{\mathrm{R}}$ for Te-terminated surface. In
Fig.~\ref{fig:kp_sdos}(b) the Fermi surface is plotted. The
topologically nontrivial hole pockets are connected by a pair of Fermi
arcs.

The $\kp$ model around K is characterized by a total Chern number of
$C=1$, respectively $C=-1$ at K'.  Hence around K and K' a quantum
Hall like surface state is expected. This is confirmed in
Fig.~\ref{fig:kp_sdos}(c) and (d), where we calculated the SDOS on a
surface orthogonal to $y$. The surface states give an excellent match
to the first principles result presented in Fig.~\ref{fig:fp}~(j-k).

We now want to take a closer look at the surface states of
TPs. Generally, one can distinguish two different scenarios.  The
first one corresponds to a pair of TPs that is very close together,
i.e. the inversion symmetry breaking is small.  Then one cannot
discriminate which of the two TPs is the source of a Fermi arc and
consequently the Fermi arcs have the characteristics of a Dirac
semimetal like in the CuPt-ordered \inassbh{} case, see
Fig.~\ref{fig:inassb}~(f)~\cite{winkler2016topological}.  In the other
scenario, one deals with isolated or a well separated pair of
TPs. Here it is more useful to consider each TP as two degenerate Weyl
nodes with opposite chirality, see Fig.~\ref{fig:fermi_simple}~(c). In
this case it is natural that each TP has two Fermi arcs connecting
with their neighbors (it is also possible that the two WPs themselves
are connected by a Fermi arc which vanishes when they form a
TP)~\cite{half_heusler}. This scenario corresponds to the TPs in
WC-like structures and in half-Heuslers.

\begin{figure}
  \includegraphics[width = \linewidth]{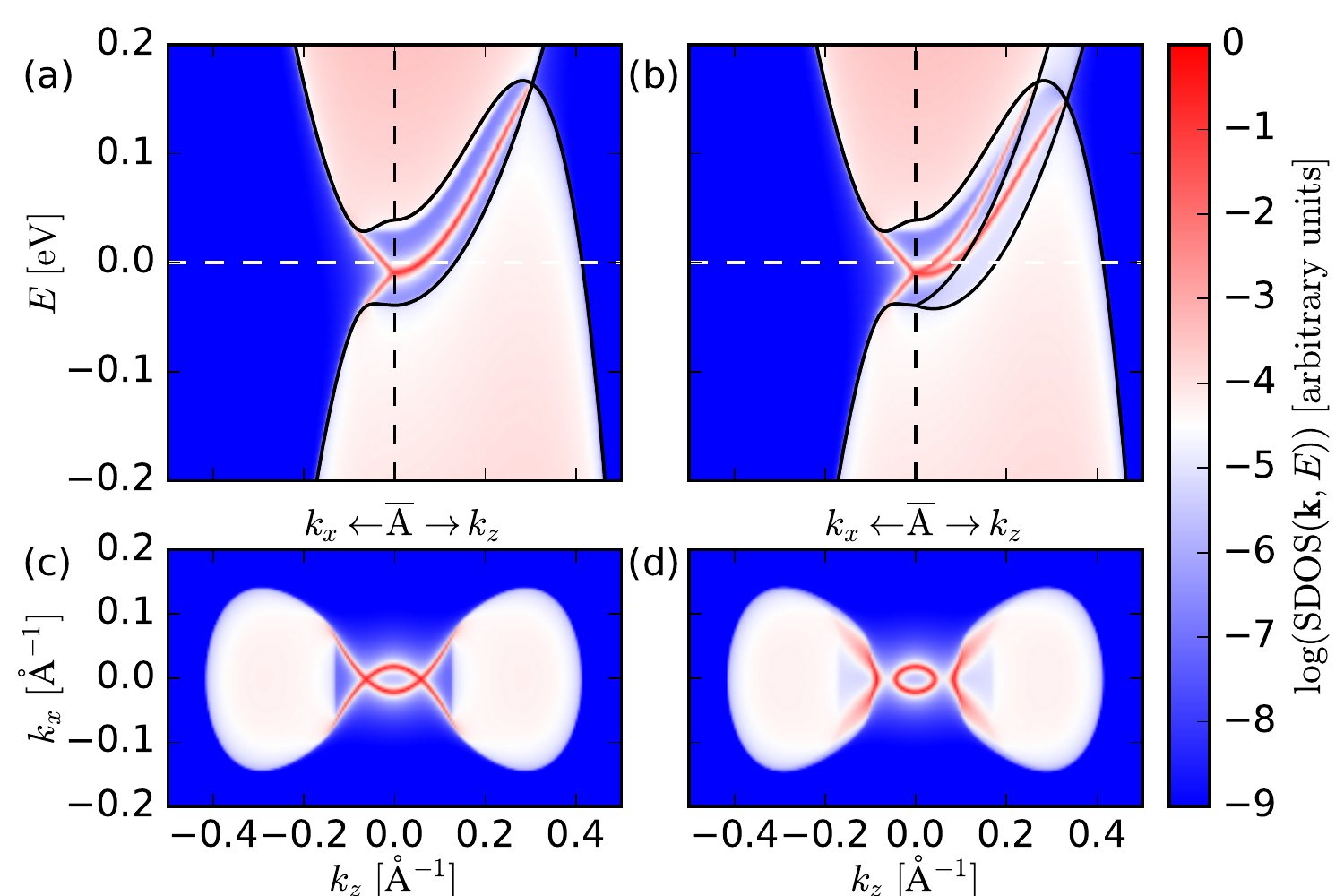}
  \caption{Surface states on the (010)-surface obtained from a $\kp$
    model for ZrTe Eq.~\eqref{eq:kpaa}. (a) ((b)) shows the surface
    density of states (the black lines show the bulk dispersion for
    $k_y=0$) and (c) ((d)) the surface Fermi surface with (without)
    inversion symmetry. Figure reused from
    Ref.~\cite{winkler2016triple}.}
  \label{fig:kp_sdos1}
\end{figure}

It is insightful to look at the evolution of Fermi arcs from Dirac
semimetal to TPTM. In a $\kp$-model for ZrTe proposed in
Ref.~\cite{winkler2016triple} one can easily tune between inversion
symmetric and asymmetric band structures.  In Fig.~\ref{fig:kp_sdos1}
we compare the surface states obtained from this model with and
without inversion symmetry. In the presence of inversion and TR
symmetry the two TPs merge into a four-fold degenerate Dirac
point. Across all energies in the gap the two hole pockets around
$\overline{\mathrm{A}}$ are connected by two Fermi arcs and the
surface state on the $k_z$-axis is two-fold degenerate. Breaking of
inversion symmetry then splits the Dirac point into two TPs. Each TP
contributes a single non-degenerate surface state. Since the two
surface states are split along the $k_z$-axis, the Fermi arcs are not
required to connect the two hole pockets. Instead, one finds a
topological-insulator-like Dirac cone around $\overline{\mathrm{A}}$
which is still protected by TR and $\sigma_h$ symmetries.

\subsection{  Magnetotransport}
The topologically non-trivial nature of Weyl semimetals reveals itself
most spectacularly in the so-called chiral anomaly of the quantum
field theory, realized by negative
magnetoresistance~\cite{adler,bell-jackiw,volovik_book,nielsen_ninomiya,qi_transport,son_spivak}.
Type-I WPs produce a gapless Landau level spectrum in a magnetic
field, whereas Type-II WPs have an anisotropic chiral
anomaly~\cite{alexey_weyl}, where the Landau level spectrum is gapless
only for certain directions of the applied magnetic field.

The magnetotransport properties of TPs also depend on the direction of
an applied magnetic field. A $C_3$ preserving magnetic field (along
the $C_3$-axis) does not gap the Landau level spectrum of a TP, but
instead each TP contributes a single chiral Landau level. However, if
the field is applied in a $C_3$-breaking direction the Landau level
spectrum becomes gapped. Such a direction dependence also occurs in
Dirac semimetals~\cite{cd3as2_landau}.

\begin{figure}
  \includegraphics[width = \linewidth]{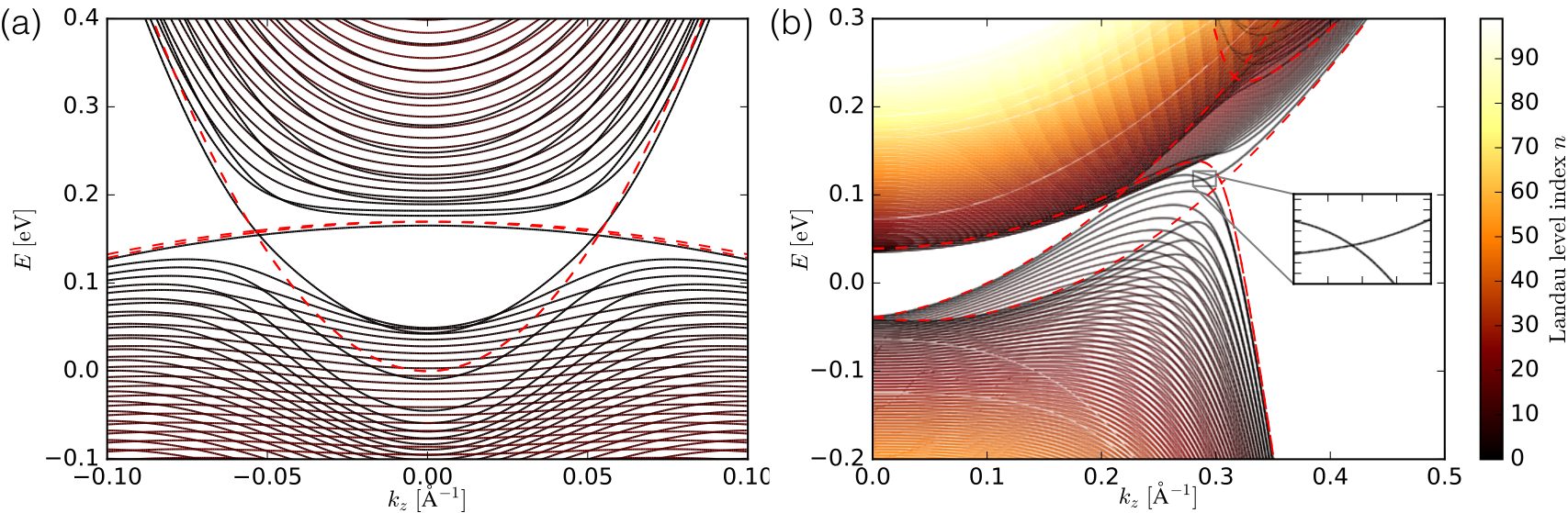}
  \caption{(a) ((b)) Landau levels in CuPt-ordered \inassbh{} (ZrTe)
    for a magnetic field of 50 (20) Tesla applied parallel to the
    $C_3$ axis using the $kp$-model of Eq.~\eqref{eq:kpb}
    (Eq.~\eqref{eq:kpa}). The dashed red lines show the bulk
    bands. Figure adapted from Refs.~\cite{winkler2016topological,
      winkler2016triple}.}
  \label{fig:landau}
\end{figure}

The Landau levels are obtained by performing a Peierls substitution of
$k_x$ and $k_y$ in the $\kp$ Hamiltonian by
$k_x = \frac{i}{\sqrt{2}l_B} (a^\dag - a)$ and
$k_y = \frac{i}{\sqrt{2}l_B} (a^\dag + a)$, with
$l_B = \sqrt{\frac{\hbar}{eB}}$ the magnetic length and $a^\dag$, $a$
the raising and lowering operators
$a^\dag |n \rangle = \sqrt{n+1} | n+1 \rangle $ and
$a |n \rangle = \sqrt{n} | n-1 \rangle $.  In Fig.~\ref{fig:landau} we
show the Landau level spectrum of CuPt-ordered \inassbh{} and
ZrTe~\cite{winkler2016topological,winkler2016triple}.  When the
magnetic field is turned on a pair of TPs turns into two crossing
chiral Landau levels with opposite chirality. Due to the gapless
Landau level spectrum strong signatures of the TPTM state are
observable in magnetotransport. Analog to the chiral anomaly one can
create a $\mathbb Z_2$ anomaly by applying parallel magnetic and
electric field~\cite{z2_anomaly}.

Magnetotransport experiments in the TPTM WC confirm the presence of
direction dependent negative magnetoresistance for parallel magnetic
and electric field~\cite{WC_magnetotransport}. Due to material growth
constraints the longitudinal magnetoresistance has been only measured
in directions orthogonal to the 001 axis. Further investigations are
required to confirm that the negative magnetoresistance stems from the
TPs.

\section{Topology of phonons in triple-point metals}
After describing the topological electronic properties of triple-point metals, we shift our focus to the topological vibrational properties of these compounds. The concept of topology from the electronic bandstructure has been successfully extended to the vibrational bandstructure of materials~\cite{KaneNature2014, Stenull_PRL2016, Roman_science2015, SusstrunkPNAS2016, Huber2016}. In recent works, topologically protected band-crossings have been reported in the phonon spectrum of solid state crystals~\cite{zhangPRL2017double, EsmannPRB2018, Li-PRB18, SinghPRM2018, xie2018phononic}. Soon after the theoretical prediction of double-Weyl phonons in the phonon spectrum of transition-metal monosilicides $M$Si ($M$=Fe, Co, Mn, Re, Ru)~\cite{zhangPRL2017double}, Miao \textit{et al.}~\cite{MiaoPRL2018} experimentally reported the first observation of double Weyl points in FeSi by means of inelastic x-ray scattering measurements. This paved the way for realization of novel phononic (bosonic) quasiparticle excitations and the associated topological properties in condensed matter systems. Several other studies focused on the topology of phonons in model systems, and predicted novel quantum phenomena arising due to the non-trivial topology of phonons, such as quantum Hall conductance, quantized phonon Berry phase, topologically protected pseudo spin-polarized interface states, phonon pseudospin Hall effect, valley effects of phonons, realization of phonon diode, and topological acoustics~\cite{LZhang_PRL2010, LZhangPRL2015, PWang_PRL2015, JiuyangLNaturePhy2016, RomainNatureComm2016, YGPengNatureComm2016, YizhouPRL2017, LiuPRB2017, YizhouL_NSR2017, YuanjunNanoLett2018, XiujuanZ_CommPhys2018}. Recently, topologically protected Weyl nodal lines have been predicted in the phonon spectrum of bulk MgB$_{2}$~\cite{xie2018phononic}. 

Analogous to the physics of the electronic bandstructure, a nontrivial band-inversion between the low frequency acoustic and high frequency optical phonon modes results in topologically protected phonon band-crossings in the momentum space. Low frequency excitations of phonons near these topological band-crossing points could yield features of novel {\it bosonic} quasiparticles, such as Dirac, Weyl, and/or triple-point. Here, one main difference from the electronic structure stems from the fact that phononic excitations have {\it bosonic} character, whereas low-energy electronic excitations are {\it fermionic}. This rules out the applicability of the Pauli's exclusion principle on phononic excitations in crystalline solids. In addition to the crystalline systems, topological bosonic quasiparticles have been realized in the photonic systems, metamaterials, and optical lattices~\cite{Raghu_PRB2008, JMei_PRB2012, KaiSun_Nature2012, Khanikaev_Natu2013, LingLu_Nature2013, Rechtsman_PRL2013,  Ling_NaturePho2014, Wen_Scireports2015, Liu_JPB_2017, FulgaPRB2018}. Moreover, S{\"u}sstrunk and Huber reported that oscillations of simple pendulum in classic mechanical systems could also host topological bosonic modes~\cite{Roman_science2015, SusstrunkPNAS2016, Huber2016}. 

Although Dirac and Weyl phonon modes have been known since the last decade~\cite{EProdan_PRL2009}, triple-point phonons joined the family of topological phonons only in 2018~\cite{Li-PRB18, SinghPRM2018}. In Refs.~\cite{Li-PRB18, SinghPRM2018}, the existence of triple-point phonons was predicted in several special type-A triple-point metals having $C_{3v}$ symmetry, such as ZrTe, HfTe, TiS, TaSb, and TaBi. Notably, these materials also host Weyl fermions and triple-point fermions in their electronic spectrum. These binary compounds belong to space group \#187 ($P\bar{6}m2$ ) having 2 atoms in the primitive cell and thus, inheriting three acoustic and three optical branches in their phonon spectrum. When the mass difference ($\Delta m  = M - m$, where $M > m$) between the constituent atoms is large, the direct frequency band gap between the optical and acoustic branches ($\Delta_g$) is finite and positive at the A(0, 0, $\frac{\pi}{c}$) point of the hexagonal Brillouin zone. With decreasing $\Delta m$, the frequency band gap decreases systematically, and acoustic and optical phonon bands invert in the frequency space for certain binary compounds having small $\Delta m$. A necessary condition for such phonon band-inversion is~\cite{SinghPRM2018}: 

\begin{equation} \label{eq:delta}
     \Delta_g = \sqrt{\frac{2\beta_{\parallel}}{m}} - \sqrt{\frac{2\beta_{\perp}}{M}}.
\end{equation}

Here, $\Delta_g$ is the direct frequency band gap at the A(0, 0, $\frac{\pi}{c}$) point of Brillouin zone, $\beta_{\parallel}$ and $\beta_{\perp}$ are in-plane and out-of-plane second-order interatomic force constants between the atoms of mass $m$ and $M$, respectively.

\begin{figure}[hb!]
 \centering
 \includegraphics[width = \linewidth]{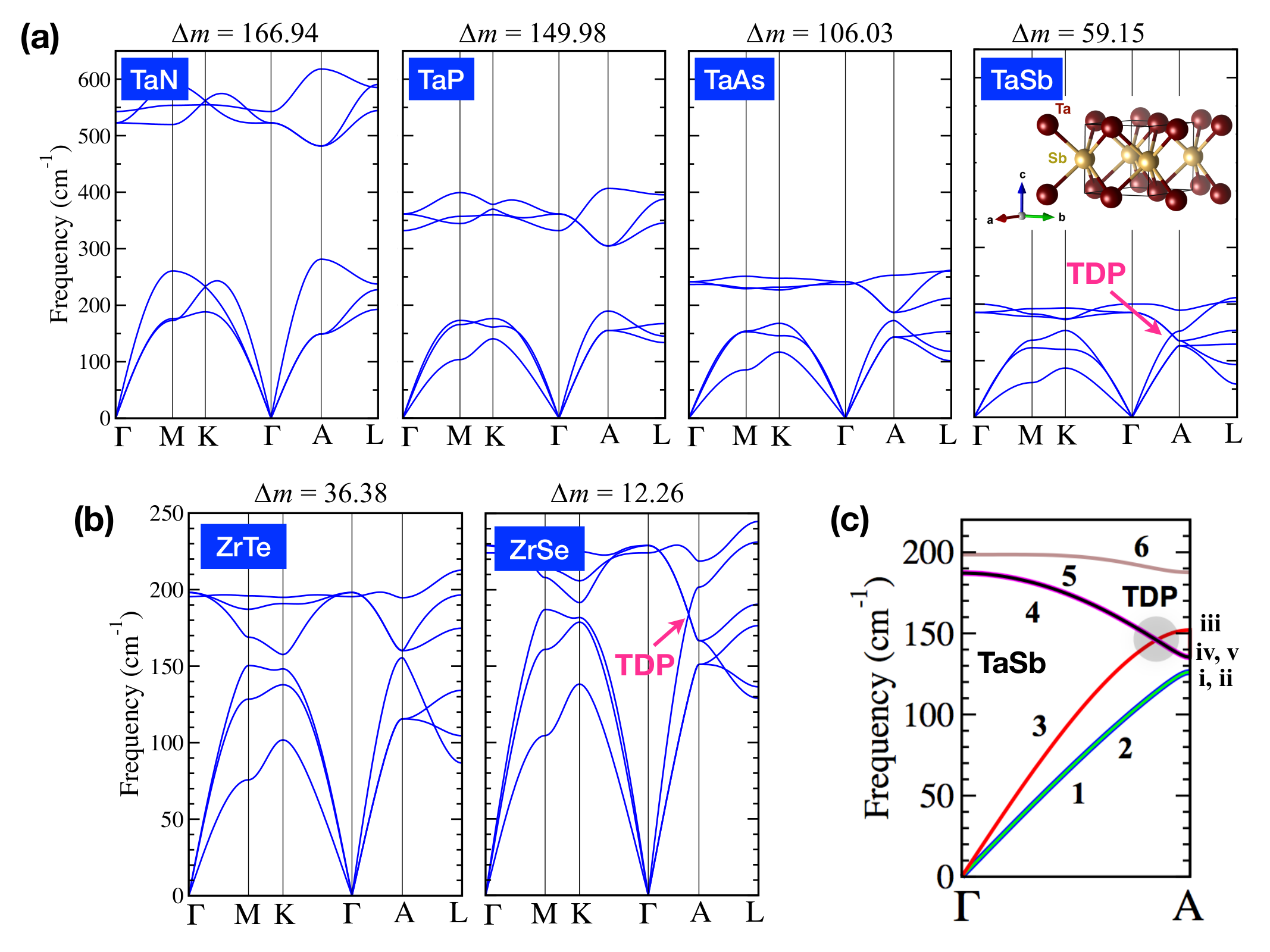}
 \caption{Phonon dispersion of (a) Ta$X$ ($X$ = N, P, As, Sb), and (b) Zr$Y$ ($Y$ = Se, Te) binaries calculated using the Density functional perturbation theory (DFPT) for supercell of size $2 \times 2 \times 2$. The mass difference $\Delta m$ is provided in $g/mol$ units. A prototype crystal structure of TaSb is given in the inset. The primitive unit cell contains one Ta and one $X$ (Sb) atoms at (2/3, 0.0, 2/3), and (0.0, 1/2, 0.0) sites, respectively. TDP is marked in the phonon spectrum of TaSb and ZrSe along $\Gamma$-A path. (c) An enlarged view of phonon dispersion in TaSb along $\Gamma$-A path. Colors and numbers depict distinct phonon modes. Arabic (Roman) numerals are used to mark phonon modes before (after) the band-inversion. Some data from Ref.~\cite{SinghPRM2018} have been reused to produce this figure. }
  \label{fig:phonon_bandinversion}
 \end{figure}

Figure~\ref{fig:phonon_bandinversion}(a) shows the calculated phonon spectra of Ta$X$ ($X$ = N, P, As, Sb) binaries having $\Delta m$ in decreasing order~\cite{SinghPRM2018}. As the  $X$ atom in Ta$X$ gets heavier, the frequency of optical phonons lowers, causing a systematic decrease in the frequency bandgap with decreasing $\Delta m$, and a phonon band-inversion takes place in TaSb along the $\Gamma$-A direction of Brillouin zone, where two degenerate optical phonon modes acquire lower frequency than a non-degenerate acoustic phonon mode. This phonon band-inversion results in a gapless triply degenerate point (TDP) in the phonon spectrum of TaSb at frequency $\sim$ 145\,cm$^{-1}$ and at $q$ = (0, 0, 0.428). This point is marked in Figure~\ref{fig:phonon_bandinversion}(a). As shown in Fig.~\ref{fig:phonon_bandinversion}(b), a similar mass induced phonon band-inversion also occurs in Zr$Y$ ($Y$ = Se, Te) family. Zr$Y$ binaries are isostructural and isoelectronic to Ta$X$~\cite{Li-PRB18}. In ZrTe, there exists a finite positive $\Delta_g$ due to the relatively large $\Delta m$. Whereas, the net frequency gap closes in ZrSe, which has relatively smaller $\Delta m$, and $\Delta_g$ becomes negative at the A-point due to the observed phonon band-inversion along $\Gamma$-A path. 

An enlarged view of the phonon dispersion along the $\Gamma$-A direction of TaSb is shown in Fig.~\ref{fig:phonon_bandinversion}(c). In this figure, the degeneracy of phonon modes and formation of a TDP can be clearly observed. Another TDP is located in other direction in the Brillouin zone at the same $C_{3v}$-symmetric line. The two degenerate optical phonon modes participating in the phonon band-crossing at TDP correspond to the in-plane ($x-y$) optical vibrations of atoms, whereas the non-degenerate acoustic mode involved in the phonon band-crossing at TDP belong to the out-of-plane ($z$) acoustic vibration of atoms. The competition between in-plane and out-of-plane interatomic force constants, and $\Delta m$ is the primary reason of the phonon-band inversion in TaSb and ZrSe~\cite{SinghPRM2018}. 

We note that the same methods~\cite{wu2017wanniertools, Z2Pack} that are used to compute the topological character of electronic band-crossings can also be employed to evaluate the topology of phonon band-crossings. However, it turns out that capturing the topology of phonon band-crossings is different from that of the electronic case. In case of non-interacting electronic systems, we analyze the symmetrized Hamiltonian of system for topological classification of the studied system. Topological classification of phononic systems can be build using the dynamical matrices  $D({\bf q})$.

\begin{figure*}[htb]
 \centering
 \includegraphics[width = \linewidth]{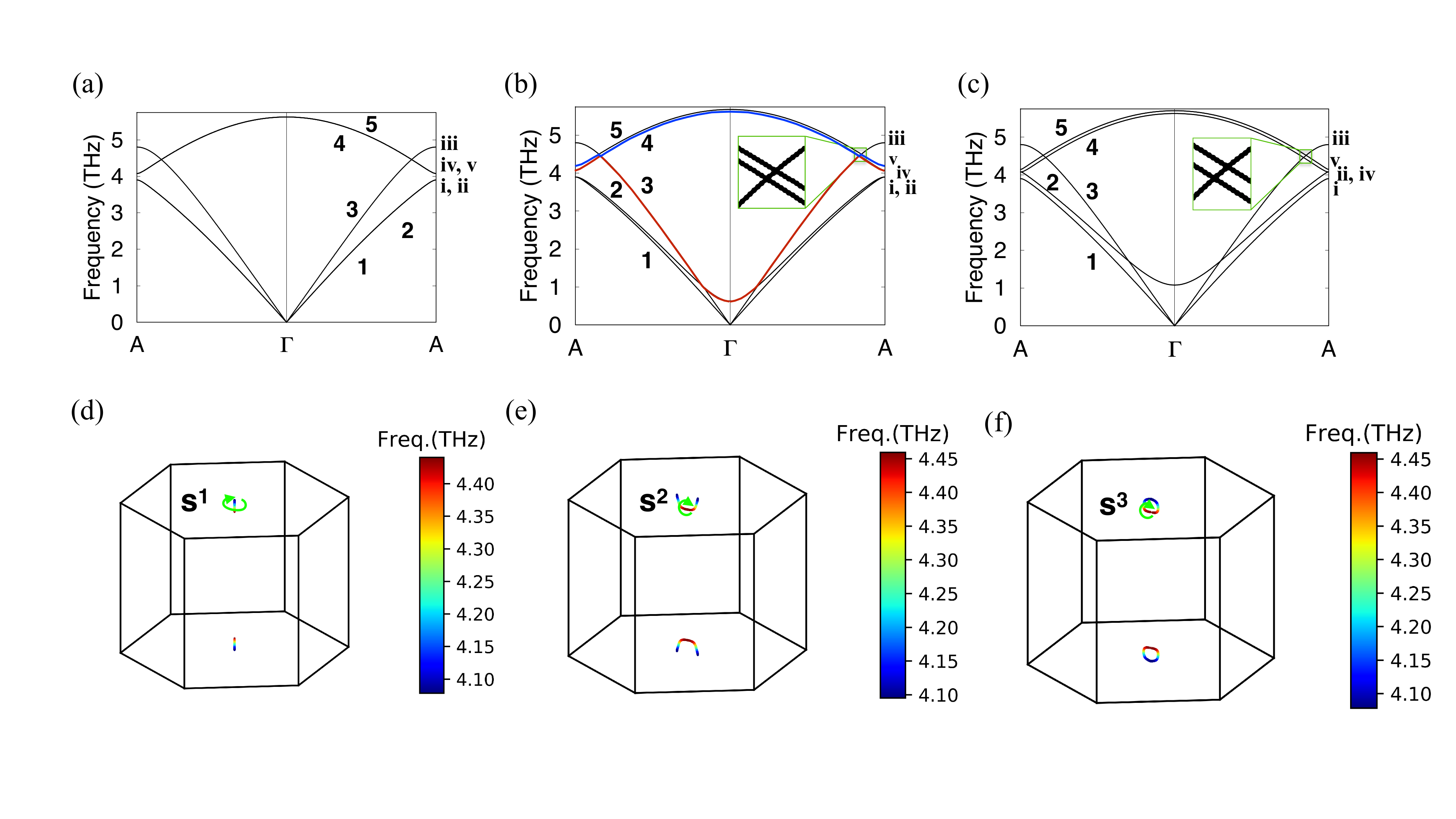}
 \caption{ (a) The phonon spectrum of TaSb along the $C_{3v}$-symmetric line of the Brillouin zone: Doubly degenerate phonon bands (4,5) and (iv,v) can be seen. (b) The phonon spectrum of TaSb with the $xx$ and $yy$ force constants in the dynamical matrix made unequal for Ta. The degeneracy of (4,5) and (iv, v) is lifted. (c) The phonon spectrum of TaSb with the $xx$ and $yy$ force constants in the dynamical matrix made unequal for Sb. The degeneracy of (4,5) and (iv, v) is lifted. (d) The non-topological \textit{open nodal line} formed in the BZ by bands iv and v of panel (a). The loop $S^1$ encircles this line, but  $\phi_B(S^1)=0$. (e) The \textit{closed nodal loop} formed by bands iv and v in the BZ for the case of panel (b) away from the $\Gamma$-A line. The contour $S^2$ links with this loop and $\phi_B(S^2)=\pi$. (f) The \textit{closed nodal loop} formed by bands iv and v in the BZ for the case of panel (c). For a contour $S^3$ linked with this loop $\phi_B(S^3)=\pi$. Figure adopted from Ref.~\cite{SinghPRM2018} with permission.}
 \label{fig:topology}
 \end{figure*}

Similar to the case of triple-points in the electronic spectrum~\cite{winkler2016triple}, gapless TDPs in the phonon spectrum are connected by a gapless nodal line in the phonon frequency space. This nodal line, shown in Fig.~\ref{fig:topology}(d), is formed by the degenerate optical phonon modes near the TDP, and it is being protected by the $C_{3v}$ rotational symmetry of the crystal. Under the $C_{3v}$  symmetry, the $xx$ and $yy$ interatomic force constants transform just like the $p_x$ and $p_y$ orbitals in $C_{3v}$ group forming a 2D irreducible representation $E$, which enforces the degeneracy of optical phonon modes along $\Gamma$-A. Computation of Berry phase ($\phi_{B}$) using the Wilson loop approach along path $S^1$, as shown in Fig.~\ref{fig:topology}(d), reveals $\phi_{B} = 0$ for the gapless nodal line connecting two TDPs~\cite{SinghPRM2018}. 

In order to reveal the topology of a TDP, one can split the two connected TDPs by breaking the $C_{3v}$ symmetry. This can be done by changing the $xx$ and $yy$ entries in the $D({\bf q})$, which is equivalent of tuning the interatomic force-constants (or equivalently, classical spring constants) between atoms along $x$ and $y$ directions~\cite{SinghPRM2018}. This trick lifts up the degeneracy of the  optical phonon modes, as shown in Fig.~\ref{fig:topology}(b-c), and enables us to probe the topological nature of TDP. It is worth noting that the aforementioned trick does not destroy the gapless nodal line, rather it results in a gapless nodal loop formed by the inverted optical and acoustic phonon bands (see Fig.~\ref{fig:topology}(e-f)). We can now define a path $S^2$ (and $S^3$) enclosing the gapless nodal loop, and compute the $\phi_{B}$ using the Wilson loop approach as demonstrated in Fig.~\ref{fig:topology}(e-f). This calculation results in $\phi_{B}$ = $\pi$ for $S^2$ and $S^3$ paths, thus confirming the topological nature of TDP~\cite{SinghPRM2018}. The mentioned technique is analogous to splitting two triple-point nodes in the electronic spectrum into four Weyl nodes by applying a Zeeman field~\cite{winkler2016triple}. 

The nontrivial topology of phonon band-crossings in bulk indicates the presence of nontrivial surface phonon states~\cite{SinghPRM2018}. In fact, Li et al. predicted unusual phonon surface states (open gapless arcs) in TiS and HfTe compounds, which are triple-point metals hosting TDPs~\cite{Li-PRB18}. 

\section{Thermoelectricity in triple-point metals}

\begin{figure*}[htb]
 \centering
 \includegraphics[width = \linewidth]{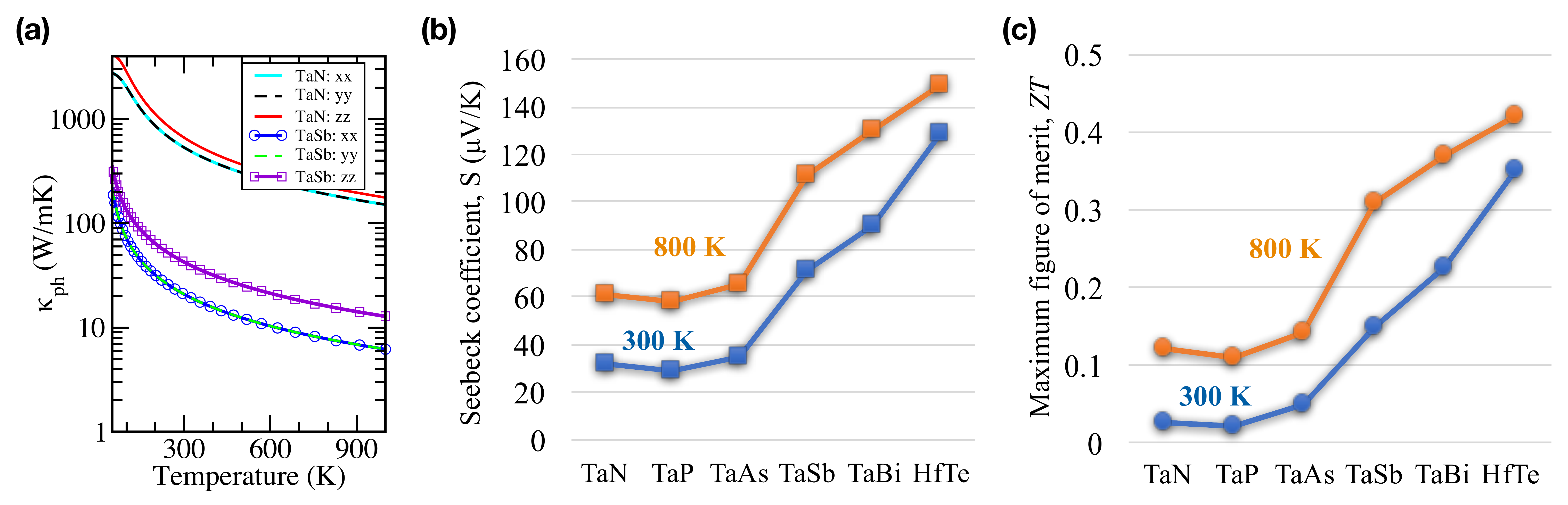}
 \caption{ (a) Comparison of the calculated lattice thermal conductivity ($\kappa_{ph}$) versus temperature data for TaN and TaSb binaries. Vertical axis is in logarithmic scale. (b) The maximum estimated thermopower or Seebeck coefficient, and (c) maximum estimated  thermoelectric figure of merit computed at 300\,K and 800\,K for electronic doping case. Data from Ref.~\cite{SinghPRM2018} is reused to produce this figure.}
  \label{fig:thermoelectric_fig}  
   \end{figure*}

Since the nontrivial topology of electronic bands is often manifested in the electronic transport measurements, the phonon topology could have observable signatures in the lattice thermal transport measurements. It has been reported that topologically protected nontrivial phonon band-crossings introduce robust phonon-phonon scattering centers that enormously suppress the lattice thermal conductivity ($\kappa_{ph}$) in triple-point metals hosting TDPs~\cite{SinghPRM2018}. In Figure~\ref{fig:thermoelectric_fig}(a), we compare the $\kappa_{ph}$ of two isostructural and isoelectronic triple-point metals-- TaN and TaSb, where the later hosts TDPs whereas the former does not. We notice that the overall $\kappa_{ph}$ in TaSb is almost three orders in magnitude smaller than that of in TaN. Such a large reduction in $\kappa_{ph}$ causes a direct benefit towards the improved thermoelectric performance of triple-point metals hosting TDPs~\cite{SinghPRM2018}.  

The thermoelectric figure of merit, $ZT$ = $\frac{S^{2}\sigma}{\kappa_{el} + \kappa_{ph}}T$, is the key indicator of the thermoelectric performance of any material. To achieve large $ZT$, one requires large thermopower or Seebeck coefficient ($S$), high electrical conductivity ($\sigma$), low electronic thermal conductivity ($\kappa_{el}$), and low lattice thermal conductivity ($\kappa_{ph}$) at a given temperature (T). Unfortunately, $\sigma$ and $\kappa_{el}$ are intrinsically coupled following the Wiedemann-Franz law, {\it i.e. } $\kappa_{el} = L_{0} T \sigma$, where $L_{0}$ is the Lorenz number~\cite{Snyder2008}. Therefore, an optimal balance is required between $\sigma$ and $\kappa_{el}$. Another key factor is the thermopower or Seebeck coefficient ($S$) which is primarily governed by the electronic electronic density of states near the Fermi-level. Generally, narrow bandgap semiconductors (with small $\kappa_{el} + \kappa_{ph}$) exhibit the best thermoelectric performance ($ZT$ $\sim$ 2-3) when slightly doped by $n-$ or $p-$type charge carriers~\cite{Snyder2008, rowe2005thermoelectrics, Zhao_NatureComm2014, ADuong_NatureComm2016, Zhao141_science1016, singh2016PCCP}. Whereas, metals are known to be poor thermoelectrics due to their large $\sigma$ and large $\kappa_{el}$ (because of no energy bandgap). Moreover, $\kappa_{ph}$ is also considerably large in most of the metals, thus, overall $ZT$ is quite low in metals ranging from $0.0001$ -- $0.001$~\cite{TerasakiPRB1997, TakahataPRB2000, OkudaPRB2001}.

The triple-point metals hosting TDPs seem to be special metals having considerably large and sizable $ZT$. The maximum estimated $ZT$ in triple-point metals hosting TDPs ranges from $\sim$0.15 (at room temperature) to $\sim$0.4 (at 800\,K)~\cite{SinghPRM2018}. Remarkably, in HfTe (which is a triple-point metal hosting TDPs) the maximum estimated $ZT$ ranges from 0.35 (at 300\,K) to 0.42 (at 800\,K) for electron doping (see Fig.~\ref{fig:thermoelectric_fig}(b-c)). Evidently, the thermoelectric performance of triple-point metals hosting TDPs is 2--3 orders in magnitude larger than that of the trivial metals~\cite{SinghPRM2018}. 

Two key factors combine to enhance the net $ZT$ in triple-point metals. First, the presence of TDP, which gives rise to reduced $\kappa_{ph}$. Second, the nontrivial electronic band-crossings near the Fermi-level, {\it i.e.} the presence of gapless Weyl/Triple-points fermions yields large $\sigma$ (unfortunately, $\kappa_{el}$ is also large) as well as enhanced  thermopower $S$. Notably, $\sigma$ lies in the same order of magnitude ($\sigma/\tau \sim 10^{20}\,1/{\si{\ohm}}ms$ at low-doping concentrations, $\tau$ being the electron-phonon relaxation time) for all the yet studied triple-point metals~\cite{SinghPRM2018}. Gapless points near the Fermi-level give rise to sharp enhancement in the electronic density of states near the Fermi-level, which increases $S$ according to the Mahan-Sofo theory~\cite{Mahan7436}. Thus, triple-point metals with TDPs inherit high $S$, high $\sigma$, and low $\kappa_{ph}$, which consequently yields relatively better thermoelectric performance in such metals. Figures~\ref{fig:thermoelectric_fig}(b-c) compare the thermopower ($S$) and thermoelectric performance ($ZT$) of selected triple-point metals with and without TDPs in their phonon spectra. For more technical details regarding the calculation of vibrational and thermoelectric properties, we recommend the reader to Ref.~\cite{SinghPRM2018}.

\section{Concluding Remarks}

We have overviewed the theoretical and experimental progress in the
field of triple-point fermions protected by symmorphic
symmetries. While the topological classification and basic band
structure of triple-points are now well understood, open challenges
lie ahead in investigating the transport and optical properties. In
particular low temperature applications could arise due to the
presence of direction dependent magnetotransport.

Experimentalists are always in need of more accessible material
candidates to confirm the manifold theoretical predictions. Recent
experimental progress has been made for MoP~\cite{MoP1, MoP2}, where
the existence of triple-points has been directly confirmed by ARPES,
and WC~\cite{WC_magnetotransport, WC_arpes}, of which the fermi arcs
and magnetotransport of triple-points have been investigated. To
improve on this studies materials with good growth characteristics and
well separated triple-points close to the Fermi level are desired. The
recently predicted half-Heusler materials seem to have very favorable
properties in this respect~\cite{half_heusler}.

Lastly, we have surveyed the topology of phonons in condensed matter systems, particularly in triple-point metals, and also discussed the numerical techniques required to evaluate the nontrivial topological nature of the triple-point nodes present in the phonon spectrum. The presence of topological phonon band-crossings usually introduce phonon-phonon scattering centers yielding low lattice thermal conductivity ({\it{phonon glass}}), whereas presence of gapless topological nodes near the Fermi-level yield good electrical conductivity and thermopower ({\it {electron crystal}}). These two factors ({\it {electron crystal}} + {\it{phonon glass}}) combine to improve the thermoelectric performance of certain triple-point metals that host TDPs in their phonon spectra. Although the thermoelectric performance of triple-point metals is not comparable to that of the best known thermoelectrics (it is obviously expected given the metallicity), triple-point metals with topological phonon band-crossings rank much better when compared to the thermoelectricity of ordinary metals. 

To conclude, the discovery and description of triple-point metals will
allow for further progress in understanding topological phenomena in
solids and identification of topological materials with potential
applications in technology. The triple-point fermion has been accepted
as an integral part of the family of topological (semi-) metals.

\section*{Acknowledgment}
  Special gratitude goes to QuanSheng Wu, who made significant contributions to the understanding and illustration of the triple-point metal physics. We would also like to thank Z. Zu, J. Li, P. Krogstrup and M. Troyer, with
  whom we collaborated on the fermionic part of our research. We acknowledge A. H. Romero and C. Yue, who made significant contribution to our study of the phonon topology.  We also thank D. Gresch,
  T. Hyart, R. Skolasinski, J. Cano, J. W. Liu, C. M. Marcus and
  M. Wimmer for useful discussions. G. W. W.  and A. A. S. were supported by Microsoft
  Research and the Swiss National Science Foundation (SNSF) through the
  Nacional Competence Centers in Research MARVEL and QSIT. A. A. S. acknowledges the support of the SNSF Professorship grant. S.S. acknowledges the support from the Dr. Mohindar S. Seehra Research Award.


\bibliographystyle{my_bibstyle_CPB_iopart-num}

\bibliography{refs}

\end{document}